\documentclass[11pt,draftcls,onecolumn]{IEEEtran}
\usepackage{amsfonts,amssymb,amsbsy,amsmath,paralist,theorem,bm,ifthen,color}
\usepackage[pdfstartview=FitH,bookmarksnumbered,unicode,bookmarksopen=true]{hyperref}
\usepackage{graphicx}
\usepackage{subfigure,relsize}
\usepackage[numbers,sort&compress,square]{natbib}
\usepackage{cases,booktabs}
\usepackage{algorithmic,algorithm}
\usepackage{array}

\usepackage[table]{xcolor}
\usepackage{enumerate}
\usepackage{epstopdf}
\usepackage{multirow}
\usepackage{mathtools}
\graphicspath{{transfigure/}}
\usepackage{xcolor}
\usepackage[numbers,sort&compress,square]{natbib}
\usepackage{caption}

\newtheorem{prop}{Proposition}

\newtheorem{theo}{Theorem}

\newcommand\Kc{\ensuremath{\mathcal{K}}}

\newcommand\Cbb{\ensuremath{{\mathbb{C}}}}

\newcommand\Ebb{\ensuremath{{\mathbb{E}}}}

\newcommand\wb{\ensuremath{{\bm w}}}

\newcommand\Hb{\ensuremath{{\bm H}}}
\newcommand\hb{\ensuremath{{\bm h}}}
\newcommand\Ab{\ensuremath{{\bf A}}}
\newcommand\ab{\ensuremath{{\bm a}}}

\newcommand\Gb{\ensuremath{{\bm G}}}
\newcommand\gb{\ensuremath{{\bm g}}}
\newcommand\Ib{\ensuremath{{\bm I}}}

\newcommand\Ub{\ensuremath{{\bf U}}}

\newcommand\vb{\ensuremath{{\bm v}}}
\newcommand\Wb{\ensuremath{{\bf W}}}

\newcommand\xib{\ensuremath{{\bm \xi}}}
\newcommand\LambdaB{\ensuremath{{\bf \Lambda}}}

\newcommand\thetab{\ensuremath{{\bm \theta}}}

\newcommand\SINR{\ensuremath{{\sf SINR}}}

\newcommand\Hf{\ensuremath{{\mathsf{H}}}}
\newcommand\Tf{\ensuremath{{\mathsf{T}}}}

\begin{document}
\title{Learning to Beamform in Heterogeneous Massive MIMO Networks}
\author{Minghe Zhu, Tsung-Hui Chang and Mingyi Hong 
\thanks{\smaller[1] Part of the work was presented in IEEE SAM 2020 \cite{zhu2020optimization}.}
\thanks{\smaller[1] M. Zhu and T.-H. Chang are with the Shenzhen Research Institute of Big Data, and School of Science and Engineering, The Chinese University of Hong Kong, Shenzhen, 518172, China (e-mail: minghezhu@link.cuhk.edu.cn,~tsunghui.chang@ieee.org).}
\thanks{\smaller[1] M. Hong is with the Department of Electrical and Computer Engineering, University of Minnesota, Twin cities, MN, USA (e-mail: mhong@umn.edu). }
}
\maketitle	

\begin{abstract}
It is well-known that the problem of finding the optimal beamformers in massive multiple-input multiple-output (MIMO) networks is challenging because of its non-convexity, and conventional optimization based algorithms suffer from high computational costs. While computationally efficient deep learning based methods have been proposed, their complexity heavily relies upon system parameters such as the number of transmit antennas, and therefore these methods typically do not generalize  well when deployed in heterogeneous scenarios where the base stations (BSs) are equipped with different numbers of transmit antennas and have different inter-BS distances. This paper proposes a novel deep learning based beamforming algorithm to address the above challenges. Specifically, we consider the weighted sum rate (WSR) maximization problem in multi-input and single-output (MISO) interference channels, and propose a deep neural network architecture by unfolding a parallel gradient projection algorithm.  Somewhat surprisingly, by leveraging the low-dimensional structures of the optimal beamforming solution, our constructed neural network can be made independent of the numbers of transmit antennas and BSs. Moreover, such a design can be further extended to a cooperative multicell network. Numerical results based on both synthetic and ray-tracing channel models show that the proposed  neural network can achieve  high  WSRs with significantly reduced runtime, while exhibiting favorable generalization capability with respect to the antenna number, BS number and the inter-BS distance.
\end{abstract}

\section{Introduction}
\vspace{-0.0cm}
\label{sec:intro}
Massive multiple-input multiple-output (MIMO) is an emerging technology that uses antenna arrays with a few hundred antennas simultaneously serving many tens of terminals in the same time-frequency resource \cite{larsson2014massive}. Multiuser beamforming techniques based on massive MIMO can provide high spectral efficiency and have been recognized as a key technology for 5G wireless networks \cite{parkvall2017nr}.
%

However, there are many challenges in searching for optimal beamforming strategies for effectively improving the performance of wireless communication systems. First and foremost, the computational complexity brought by significantly increased number of base stations (BSs) and transmit antennas is too high to fulfill the latency requirement of 5G applications. 
Moreover, relying on the massive MIMO and millimeter wave (mmWave) communication technologies \cite{ge20165g}, ultra-dense cellular networks composed of a large number of small cells and macrocells are expected to be deployed heterogeneously in cellular scenarios. There will be different numbers of access points (APs) installed inside the building or BSs located outdoors which are equipped with different numbers of antennas to deal with the complex communication environments. 
Beamforming optimization for these dense and heterogeneous networks have to jointly consider different network sizes and antenna configurations, making it much harder to search for a proper beamforming solution.  
Therefore, a multiuser beamforming method that is computationally scalable in heterogeneous networks, regardless of the network size, antenna number or user number, is highly desired.


\vspace{-0.3cm}
\subsection{Related Work}

Beamforming optimization has been an active research area in the past two decades \cite{gershman2010convex}. The power minimization based beamforming problems can be well-solved \cite{gershman2010convex} or well-approximated by various convex optimization techniques \cite{ma2010semidefinite}.
On the contrary, the weighted sum rate maximization (WSRM) based beamforming problem is difficult to solve and in fact NP-hard in general \cite{luo2008dynamic,liu2010coordinated}. Many suboptimal but computationally efficient beamforming algorithms have been proposed in the literature.
For example, the paper \cite{wiesel2008zero} 
proposed the zero-forcing based beamforming based on the generalized matrix inverse theory. 
Approximation algorithms based on successive convex approximation techniques are proposed in \cite{papandriopoulos2009scale,shen2014wireless} for efficient resource allocation and multiuser beamforming optimization.
The inexact cyclic coordinate descent algorithm proposed in \cite{liu2010coordinated} relies on the block coordinate descent (BCD) and gradient projection (GP), and can achieve good performance with a low complexity. The celebrated weighted minimum mean square error (WMMSE) algorithm proposed in \cite{christensen2008weighted,shi2011iteratively} is based on the equivalence between signal-to-interference-plus-noise ratio (SINR) and MSE, which then is solved by the BCD method. The WMMSE algorithm provides the state-of-the-art performance and therefore is widely benchmarked in the literature. However, all these existing algorithms are iterative in nature, and their complexities quickly increase with the antenna number and network size.




In recent years, machine learning based approaches that depend on the deep neural network (DNN) have been considered in a range of wireless communication applications \cite{o2017deep}.
%
For instance, in \cite{cui2019spatial}, a parallel structured convolutional neural network (CNN) is trained with only geographical location information of transmitters and receivers to learn the optimal scheduling in dense device-to-device wireless networks. 
A black-box DNN is trained to approximate the WMMSE algorithm to learn the optimal power control strategy for WSRM in the interference channel \cite{sun2018learning,liang2019towards}.  The paper \cite{nasir2019multi} demonstrates the potential of multi-agent deep reinforcement learning techniques to develop a dynamic transmit power allocation scheme in wireless communication systems. 

DNN based beamforming schemes have also been proposed for alleviating the computational issues faced in massive MIMO communications.
For example, by considering the multiple-input single-output (MISO) broadcast channel, the work \cite{xia2019deep} proposed a beamforming neural network (BNN) that learns virtual uplink power variables based on the well-known uplink-downlink duality \cite{gershman2010convex} for the power minimization problem and the SINR balancing problem. For the WSRM problem, they proposed a BNN to learn the power variables and Lagrange dual variables based on the optimal beamforming structure.
%
%
By considering a coordinated beamforming scenario with multiple BSs serving one receiver, the work \cite{alkhateeb2018deep} proposed a black-box DNN to learn the radio-frequency (RF) downlink beamformers directly from the signals received 
at the distributed BSs during the uplink transmission.
%

Different from the black-box DNN approach, the deep unfolding technique \cite{balatsoukas2019deep,hershey2014deep} 
can build a learning network based on approximating a known iterative algorithm with finite iterations.  
{For example, the works \cite{samuel2019learning}, \cite{un2019deep} and \cite{wei2020learned} respectively unfold the GP algorithm, alternating direction method of multipliers (ADMM) and gradicent descent algorithm to build learning networks for MIMO detection.} 
%
For a single-cell multiuser beamforming problem, the authors in \cite{hu2020iterative} proposed a learning network by unfolding the WMMSE algorithm.
To overcome the difficulty of matrix inversion involved in the WMMSE algorithm, they approximate the matrix inversion by its  first-order Taylor expansion. Another recent work \cite{pellaco2020deep} considered to unfold the WMMSE algorithm to solve the coordinated beamforming problem in MISO interference channels. They chose to avoid the matrix inversion by unfolding a GP algorithm to solve the beamforming subproblem. While these existing works have shown good performance, their learned networks cannot be easily used to optimize a new scenario in which network parameters such as the number of antennas, the number of BSs or the network size are different from the scenarios when the neural network is trained. This shortcoming makes the current designs not suitable for heterogeneous wireless environments.


%
%

\vspace{-0.3cm}
\subsection{Contributions}

\begin{figure}
	\centering
	\subfigure[]{
		\begin{minipage}{0.45\textwidth}
			\centering
			\includegraphics[width=\textwidth]{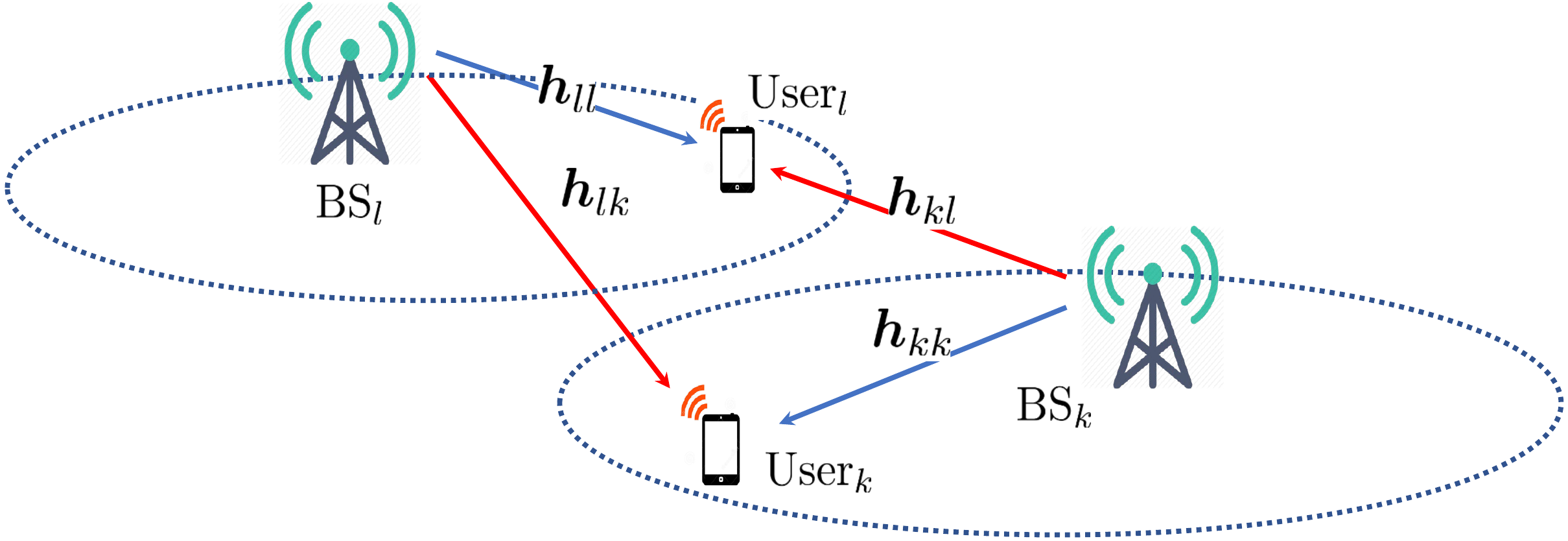}
			\label{fig:fig1}	
	\end{minipage}}\vspace{-0.1cm}
	\subfigure[]{
		\begin{minipage}{0.45\textwidth}
			\centering
			\includegraphics[width=\textwidth]{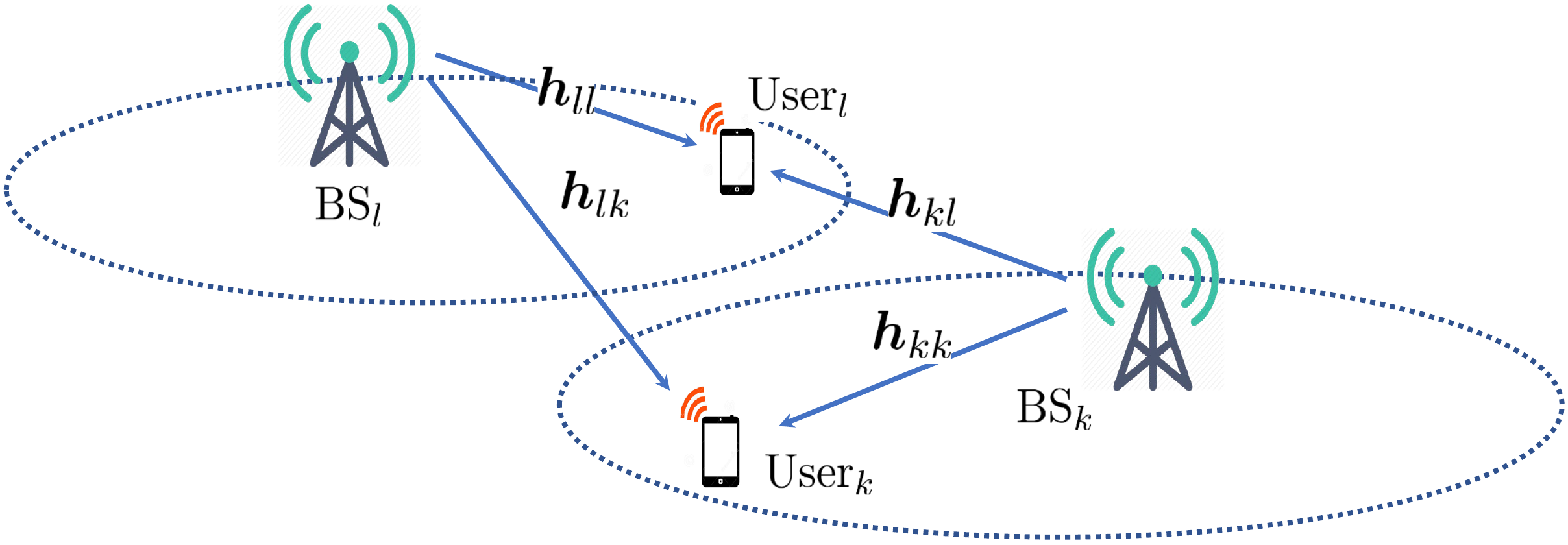}
			\label{fig:fig2}	
	\end{minipage}}\vspace{-0.1cm}
	\caption{\footnotesize (a) The MISO interference channel \cite{liu2010coordinated}, and (b) the cooperative multicell network \cite{larsson2008competition}. {The blue arrows represent information signals and the red arrows represent the inter-cell interference}. 
	}
	\vspace{-0.4cm}
	\label{fig:system model}
\end{figure}

In this paper, we consider learning-based beamforming designs for WSRM in the MISO interference channels as well as the cooperative multicell networks (see Fig. \ref{fig:system model}).
We propose a beamforming learning network by unfolding the simple parallel GP (PGP) algorithm \cite{bertsekas1999nonlinear}. 
The proposed learning network has a recurrent neural network (RNN) structure and is referred to as ``RNN-PGP".
The first key advantage of the proposed RNN-PGP is that it has a parallel structure and the deployed neural network is identical for all BSs. 
The second advantage is that the complexity of the neural network can be made {\it independent} of both the number of BSs and the number of BS antennas (which is usually large in massive MIMO communications), and thus the dimension of learnable parameters does not increase with the network size and antenna size.
Consequently, the proposed RNN-PGP has the third advantage that it has good generalization capability with respect to the cell radius, the number of antennas, and the number of BSs, which means that the proposed RNN-PGP  can be easily deployed in heterogeneous networks where the BSs may have different number of antennas and non-uniform inter-BS distances. 
%
%
Our specific technical contributions are summarized as follows.

\begin{enumerate}
	\item We propose a RNN-PGP beamforming learning network by unfolding the PGP algorithm, which is a simple but effective method to handle the WSRM problem. 
	In order to accelerate the algorithm convergence, we employ a multi-layer perception (MLP) to predict the gradient vector with respect to the beamforming vectors. By showing that the gradient vector lies in a low-dimensional subspace, the MLP simply learns the coefficients required to construct the gradient vector. Since the MLP is identical across the RNN iterations and for all BSs, the parameter space to be learned can be small.
	%
	
	\item We make two key steps to improve the generalization capability of RNN-PGP with respect to the number of transmit antennas and the number of BSs. Firstly, the low-dimensional structure of the optimal beamforming solution is exploited to transform the target WSRM problem into a dimension-reduced problem. This ensures that the proposed RNN-PGP actually solves a problem whose dimension is independent of the number of transmit antennas. Secondly, instead of considering the interference from the whole network, the MLP only considers the signals and interference that are sufficiently strong to predict the beamforming gradient vector. 
	This makes the MLP dimension independent of the whole network size, and thus the RNN-PGP can be robust against the number of BSs.
	
	%
	%
	
	\item The design of the RNN-PGP is further extended to the cooperative multicell beamforming problem, which is a joint transmission scheme with BS cooperation. The performance of the proposed RNN-PGP is examined by both synthetic channel dataset and  ray-tracing based DeepMIMO dataset \cite{alkhateeb2019deepmimo}. The results show that the proposed RNN-PGP can well approximate the WMMSE solution. More importantly, the RNN-PGP shows promising generalization capability with respect to the number of BSs, number of antennas and inter-BS distance.
\end{enumerate}


{\bf Synopsis:}
Section \ref{sec: system model} presents the system model of the MISO interference channel and formulates the WSRM problem. The existing beamforming algorithms are also briefly reviewed.
Section \ref{sec: PGP Inspired RNN} presents the main design of the proposed RNN-PGP, including introduction of the approach to improving its generalization capability.
%
%
In Section \ref{sec:other applications}, we extend our RNN-PGP to solve the more complex cooperative multicell beamforming problem. 
The simulation results are given in Section \ref{sec:simulation} and the paper is concluded in Section \ref{sec: conclusion}.

{\bf Notations:} Column vectors and matrices are respectively written in boldfaced lower-case and upper-case letters, e.g., $\ab$ and $\Ab$, respectively. 
The superscripts $(\cdot)^{\Tf}$, $(\cdot)^{*}$ and $(\cdot)^{\Hf}$ represent the transpose, conjugate and hermitian transpose respectively.
$\Ib_K$ is the $K \times K$ identity matrix; $\|\ab\|$ denotes the Euclidean norm of vector $\ab$. 
$\Im(\cdot)$ and $\Re(\cdot)$ represent the imaginary and real part of a complex value respectively.
$\{a_{jk}\}$ denotes the set of all $a_{jk}$ with subscripts $j,k$ covering all the admissible intergers, $\{a_{jk}\}_k$ denotes the set of all $a_{jk}$ with the first subscript equal to $j$. 


\vspace{-0.2cm}
\section{Signal Model and Problem Formulation}\label{sec: system model}
In this section, we build the signal model of the MISO interference channel, and formulate the beamforming problem for maximizing the network weighted sum rate. Then, we briefly review some existing algorithms for solving the weighted sum-rate maximization (WSRM) problem.

As shown in Fig. \ref{fig:fig1}, the downlink multi-user MISO interference channel has $K$ BSs serving $K$ respective user equipment (UE) at the same time and over the same spectrum.
Each BS is equipped with $N_t$ transmit antennas, and each UE has only one receive antenna. 
Let $s_k\in\Cbb$, $\Ebb[\vert s_k\vert^2] = 1$, be the information signal for $\text{UE}_k$, and $\vb_k\in \Cbb^{N_t}$ denote the beamforming vector used by $\text{BS}_k$, for all $k\in \Kc\coloneqq \{1,\ldots,K\}$.
Moreover, denote $\hb_{jk}\in\Cbb^{N_t}$ as the channel between $\text{BS}_j$ and$\text{UE}_k$. Then, the signal received by $\text{UE}_k$ is given by
\begin{align}
	y_k = \hb^\Hf_{kk}\vb_ks_k + \sum_{j=1,j\neq k}^{K}\hb^\Hf_{jk}\vb_js_j + n_k, ~k\in \Kc,
\end{align}
where $n_{k} \in \Cbb $ is the additive white Gaussian noise (AWGN) with zero mean and variance $\sigma_{k}^2$, i.e., $n_k \sim\mathcal{CN}(0,\sigma_{k}^2)$. 
It is assumed that the signals $s_k,~k\in \Kc$, are statistically independent from each other and from the AWGN. 
Thus, the SINR for each ${\rm UE}_k$ can be written as
\begin{align}\label{sinr}
	\SINR_k(\{\vb_{k}\},\{\hb_{jk}\}_j) = \frac{|\hb_{kk}^\Hf\vb_k|^2}{\sum_{j\neq k}|\hb_{jk}^\Hf\vb_j|^2 + \sigma_{k}^2},~ k\in\Kc.
\end{align}
Assume that the BSs have perfect channel state information (CSI). The downlink transmission rate of link $k$ can be expressed as
\begin{equation}\label{spectral efficiency}
R_k(\{\vb_{k}\},\{\hb_{jk}\}_j) = \log_2\left(1 + \SINR_k(\{\vb_{k}\},\{\hb_{jk}\}_j)\right).
\end{equation}

We are interested in designing the beamformers so that the network throughput is maximized. Specifically, the WSRM problem is formulated as
\begin{equation}\label{eqn: miso sum-rate maximization}
\begin{aligned}
\max_{\substack{\vb_k \in \Cbb^{N_t}, k\in\Kc}}~ & R(\{\vb_{k}\},\{\hb_{jk}\}) \\
{\rm s.t.}~ & \lVert \vb_{k}\rVert^2 \leq P_k, k\in\Kc, 
\end{aligned}
\end{equation}
where 
\begin{equation}\label{eqn: sum rate}
R(\{\vb_{k}\},\{\hb_{jk}\})= \sum_{k=1}^{K} \alpha_k \cdot R_k(\{\vb_{k}\},\{\hb_{jk}\}_j) ,
\end{equation}
in which $\alpha_k\geq 0$ is a non-negative weighting coefficient of link $k$, and $P_k$ denotes the maximum power budget of $\text{BS}_k$. 
Hence, the WSRM problem in the form of \eqref{eqn: miso sum-rate maximization} has to be solved before the BSs transmit signals to their receivers. 
However,  \eqref{eqn: miso sum-rate maximization} is a non-convex problem, and it has been shown to be NP-hard in general \cite{luo2008dynamic,liu2010coordinated}. In view of this, suboptimal but computationally efficient algorithms have been proposed for problem \eqref{eqn: miso sum-rate maximization}.  Next, we review the WMMSE algorithm \cite{shi2011iteratively}, the GP algorithm \cite{bertsekas1999nonlinear,liu2010coordinated}, and the POA algorithm \cite{tuy2000monotonic}.

\subsection{WMMSE Algorithm}
The WMMSE algorithm \cite{shi2011iteratively} is one of the most popular algorithms for handling the WSRM problem in \eqref{eqn: miso sum-rate maximization}. 
It reformulates \eqref{eqn: miso sum-rate maximization} as an equivalent weighted MSE minimization problem by the MMSE-SINR equality \cite{verdu1998multiuser}, followed by solving the problem with the BCD method \cite{bertsekas1989parallel}. The iterative steps of WMMSE are given by: for iteration $r =1,\ldots,$ perform

\vspace{-0.3cm}
{\small\begin{subequations}\label{Alg1}
		\begin{align}
			u_k^{r} &= \left(\sum\limits_{j=1}^K|\hb_{jk}^\Hf\vb_j^{r-1}|^2 + \sigma_{k}^2\right)^{-1}\hspace{-0.1cm}\hb_{kk}^\Hf\vb_{k}^{r-1}, \label{alg1: u}\\
			w_k^{r} &= \left(1 - u^{r}_k\hb_{kk}^\Hf\vb_{k}^{r-1}\right)^{-1}, \label{alg1: w}\\
			\vb_k^{r} &= \alpha_{k}\left(\sum\limits_{j=1}^K\alpha_{j}|u_j^{r}|^2w_j^{r}\hb_{kj}\hb_{kj}^\Tf + \mu_k^*\Ib_N\right)^{-1}\hspace{-0.35cm} u_k^{r}w_k^{r}\hb_{kk},\label{alg1: v}
		\end{align}
\end{subequations}}\vspace{-0.3cm}

\noindent for all $k\in\Kc$. In \eqref{alg1: v}, $\mu_k^*$ is an optimal dual variable associated with the power budget constraint \cite{shi2011iteratively}.

It is worth mentioning that the WMMSE algorithm developed in \cite{shi2011iteratively} was for a more general multi-user scenario where one BS could serve multiple UEs, and it can be extended to network MIMO scenarios \cite{larsson2014massive}.
Theoretically, it has been shown that the WMMSE algorithm can converge to a stationary solution of \eqref{eqn: miso sum-rate maximization}, and practically performs well.

\vspace{-0.3cm}
\subsection{Gradient ascent based Algorithm}

Since the power budget constraints of problem \eqref{eqn: miso sum-rate maximization} have a simple structure, gradient ascent based methods, such as the gradient projection (GP) method \cite{bertsekas1999nonlinear}, can be applied. For example, the inexact cyclic coordinate descent (ICCD) algorithm proposed in \cite{liu2010coordinated} deals with problem \eqref{eqn: miso sum-rate maximization} by applying gradient projection update for each beamformer $\vb_k$ in a sequential and cyclic fashion. Specifically, the ICCD algorithm has the following steps: for iteration $r =1,2,\ldots,$ perform for $k=1,\ldots,K$ sequentially
\begin{align}\label{Alg2}
	\tilde{\vb}_{k}^{r} & = \vb_{k}^{r-1} + s_k^{r-1}\nabla_{\vb_k} R(\{\vb_{j}^{r}\}_{j< k}, \{\vb_{j}^{r-1}\}_{j\geq k}, \{\hb_{jk}\}), \notag \\
	\vb_{k}^{r} &=\frac{\tilde{\vb}_{k}^{r}}{\max\{\|\tilde{\vb}_{k}^{r}\|/\sqrt{P_k},1\}}, 
\end{align}
where $s_k^r>0$ is the step size, and the step in \eqref{Alg2} is projection onto the power budget constraint in \eqref{eqn: miso sum-rate maximization}. A key advantage of the gradient ascent based methods is that they involve simple computation steps. However, when comparing to the WMMSE algorithm,  the gradient ascent based methods usually require a larger number of iterations to converge to a solution as good as the WMMSE solution.

\vspace{-0.3cm}
\subsection{POA Algorithm}
The POA algorithm {\cite{utschick2011monotonic}} 
is a monotonic optimization method which can globally solve problem \eqref{eqn: miso sum-rate maximization}. Specifically,  in the POA algorithm, a sequence of surrogate problems are systematically constructed, whose feasible set contains the feasible set of the original problem. The constructed feasible set will shrink iteratively and converge to the true feasible set of the original problem \cite{li2015multicell}, while the objective values of the constructed problems will converge to the true optimal value from above asymptotically. Therefore, the POA algorithm provides an upper bound solution for the original optimization problem \eqref{eqn: miso sum-rate maximization}. However, the POA algorithm suffers from very high computation complexity, making it impractical to be used in real-time scenarios. 

It is worth noting that all of the above algorithms are iterative in their original form.
Therefore, all of them suffer from significant computational delays especially for massive MIMO scenarios where the numbers of cells and transmit antennas are large. To alleviate the computation issues, researchers have proposed the use of DNN \cite{alkhateeb2018deep} and the deep unfolding techniques \cite{balatsoukas2019deep} for approximating the WMMSE beamforming solution in a computationally efficient fashion. While one of the challenges of implementing the WMMSE algorithm by a deep-unfolding neural network lies in the matrix inversion in \eqref{alg1: v}, the recent work \cite{pellaco2020deep} proposed to employ a GP method to handle the update of $\vb_k^r$ in order to avoid explicitly computing the matrix inversion. This results in a double-loop deep-unfolding structure. A related work \cite{hu2020iterative} also proposed a deep unfolding based network for the WMMSE beamforming solution, where the authors approximate the matrix inversion by its first-order Taylor expansion.
However, both beamforming designs in \cite{pellaco2020deep, hu2020iterative} have limited generalization capability with respect to the number of transmit antennas or network size. In particular, the learning networks have to be retrained whenever they are deployed in a scenario with different numbers of BSs and transmit antennas.

{In the next section, we present a new beamforming learning network, which is designed to improve the generalization capability of the existing DNN based approaches.} 

\vspace{-0.0cm}
\section{Proposed Beamforming Learning Network}\label{sec: PGP Inspired RNN}
In this section, we first utilize the low-dimensional structure of the beamforming solution of problem \eqref{eqn: miso sum-rate maximization} to transform the problem into an equivalent problem with reduced dimension.
Then, by unfolding the PGP method, we develop the proposed beamforming learning network, and present approaches to enhance its generalization capability with respect to the number of transmit antennas and the network size.

\vspace{-0.3cm}
\subsection{Problem Dimension Reduction}
{Since in massive MIMO scenarios the number of transmit antennas $N_t$ is large, it is desirable to avoid  handling problem  \eqref{eqn: miso sum-rate maximization}  in its original form.}
It has been shown in \cite[Proposition 1]{jorswieck2008complete} that the optimal beamforming vectors actually have a low-dimensional structure, as stated below.
%
\begin{prop}\cite[Proposition 1]{jorswieck2008complete}\label{beamform linear structure}
	Suppose that $N_t\geq K$, and that $\{\hb_{kj}\}_j$ are linearly independent and satisfy
	\begin{equation*}
		\hb_{kj}^\Hf\hb_{kj'} \neq 0,~ \forall j,j'\in\Kc,j\neq j'.
	\end{equation*}
	Then, if $\vb_k$ is a beamforming vector that corresponds to a rate point on the Pareto boundary, there exist complex numbers $\{\xi_{kj}\}_{j=1}^K$ such that
	\begin{equation}
	\vb_{k} = \sum_{j=1}^{K}\xi_{kj}\hb_{kj} ,~
	\lVert\vb_{k}\rVert^2 = P_k.
	\end{equation}
\end{prop} 

By this property, the beamforming vector for each $\text{BS}_k$ lies in the low-dimensional subspace spanned by the channel vectors $\{\hb_{kj}\}_j$. Let 
$\Hb_k =\begin{bmatrix} \hb_{k1},\ldots,\hb_{kK} \end{bmatrix}	\in \Cbb^{N_t\times K}$ and 
$ \xib_k=\begin{bmatrix} \xi_{k1} ,\ldots\,\xi_{kK} \end{bmatrix}^\top \in \Cbb^K$. We can let $\vb_k =\Hb_k \xib_k$ for all $k\in \Kc$, and rewrite 
problem \eqref{eqn: miso sum-rate maximization} as

\vspace{-0.3cm}
{\small \begin{align}\label{eqn: channel combinationed wsm}
		\!\!\!\!\max_{\substack{\xib_k\in\Cbb^K, k\in \Kc}}~ & \sum_{k=1}^{K}\alpha_{k}\log_2\left(1 + \frac{\left|\hb_{kk}^\Hf\Hb_k\xib_k\right|^2}{\sum_{j\neq k}\left|\hb_{jk}^\Hf\Hb_j\xib_j\right|^2 + \sigma_k^2}\right) \notag \\
		{\rm s.t.}~ & \|\Hb_k\xib_k\|^2 \leq P_k, k\in\Kc.
\end{align}}\vspace{-0.3cm}


To avoid handling the ellipsoid constraint $\|\Hb_k\xib_k\|^2 \leq P_k$, we consider the eigen-decomposition of $$\Hb_k^\Hf\Hb_k=\Ub_k\LambdaB_k\Ub_k^\Hf,$$ 
where $\Ub_k \in \mathbb{C}^{K \times K}$ is the unitary eigen-matrix and $\LambdaB_k  \in \mathbb{R}^{K \times K}$ is a diagonal eigenvalue matrix. By letting $\wb_k = \LambdaB_k^{1/2}\Ub_k^\Hf\xib_k$ and $\gb_{jk} = \LambdaB_j^{-1/2}\Ub_j^\Hf\Hb_j^\Hf\hb_{jk} $, we write \eqref{eqn: channel combinationed wsm} as

\vspace{-0.3cm}
{\small	\begin{align} \label{eqn: simplified sum-rate max}
		\max_{ \substack{\substack{\wb_k\in\Cbb^K, k\in \Kc}}}~ & \sum_{k=1}^{K}\alpha_{k}\log_2\left(1 + \frac{\left|\gb_{kk}^\Hf\wb_k\right|^2}{\sum_{j\neq k}\left|\gb_{jk}^\Hf\wb_j\right|^2 + \sigma_k^2}\right) \notag \\
		{\rm s.t.}~ & \lVert\wb_k\rVert^2 \leq P_k. ~k\in\mathcal{K}. 
\end{align}}\vspace{-0.4cm}

By comparing \eqref{eqn: simplified sum-rate max} with \eqref{eqn: miso sum-rate maximization}, the number of unknown parameters are reduced from $O(KN_t)$ to $O(K^2)$. Therefore, when $N_t \gg K$, it will be beneficial to deal with 
the dimension-reduced problem \eqref{eqn: simplified sum-rate max}. {In addition, problem \eqref{eqn: simplified sum-rate max} is independent of $N_t$, and therefore a learning network based on \eqref{eqn: simplified sum-rate max}  inherently has a good generalization capability with respect to the number of transmit antennas.}

%
\begin{figure*}[t]
	\centering
	\includegraphics[width=0.8\textwidth]{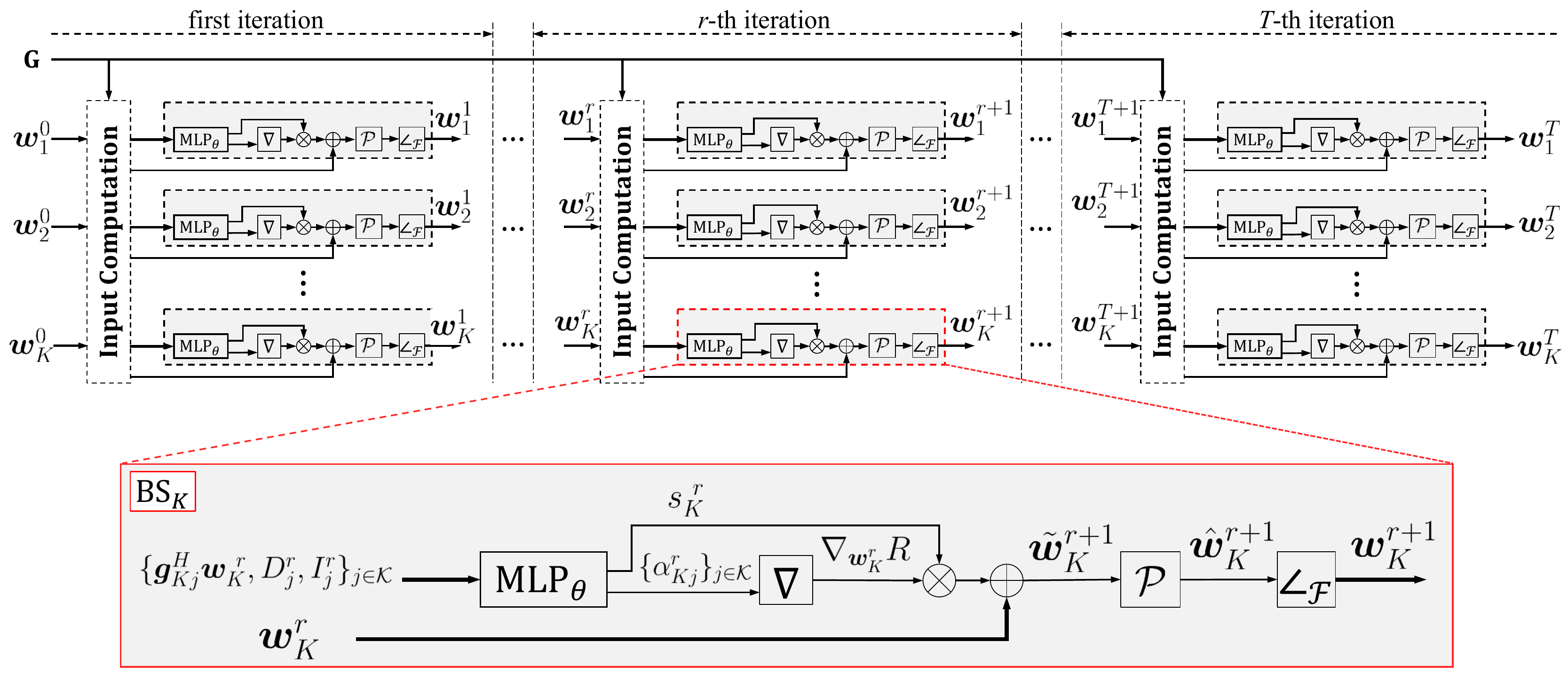}
	\vspace{-0.0cm}
	\caption{\footnotesize Diagram of the proposed RNN-PGP network for WSRM problem \eqref{eqn: simplified sum-rate max}, where $\Gb=\{\gb_{jk}\}_{j,k\in\mathcal{K}}$ contains all the transformed channel information. The subscript $(\cdot)^r_k$ refers to the parameter for the $k$-th BS in the $r$-th iteration. $\nabla$, $\mathcal{P}$ and $\angle_{\mathcal{F}}$ indicate the gradient, projection and phase rotation respectively.}
	\label{fig:pgddnn}
	\vspace{-0.2cm}
\end{figure*}

\subsection{PGP Inspired Learning Network}\label{subsec: 2-stage approx}
While the WMMSE algorithm can provide state-of-the-art performance, the matrix inversion structure of the updating rule makes it  difficult to build a learning network to learn the beamforming solution. In this section, in view of the separable power constraint, we consider the simple PGP method {\cite{bertsekas1999nonlinear}} to handle problem \eqref{eqn: simplified sum-rate max}.  Moreover, based on the deep unfolding technique, we show how an effective and computationally efficient beamforming learning network can be built.

The PGP method applied to \eqref{eqn: simplified sum-rate max} entails the following iterative steps: for iteration $r =1,\ldots,$ perform for $k=1,\ldots,K$ in parallel
\begin{align}\label{Alg3}
	\tilde{\wb}_{k}^{r} & = \wb_{k}^{r-1} + s_k^{r-1}\nabla_{\wb_k} R( \{\wb_{j}^{r-1}\}, \Gb),  \\
	\wb_{k}^{r} &=\frac{\tilde{\wb}_{k}^{r}}{\max\{\|\tilde{\wb}_{k}^{r}\|/\sqrt{P_k},1\}}, \label{eqn: projection 2}
\end{align} where $\Gb=\{\gb_{jk}\}_{j,k\in\mathcal{K}}$ contains all the transformed channel information.
According to the deep unfolding idea, one can build an RNN with a finite iteration number to imitate the iterative updates of the PGP method \eqref{Alg3}-\eqref{eqn: projection 2}. In the existing works such as \cite{pellaco2020deep}, only the step sizes $\{s_k^{r-1}\}$ are set to the learnable parameters. Here, we attempt to learn both the step size $s_k^{r-1}$ and the gradient vector $\nabla_{\wb_k} R( \{\wb_{j}^{r-1}\}, \Gb\})$, aiming to find good ascent directions that may expedite the algorithm convergence.

It is interesting to note that the gradient vector $\nabla_{\wb_k} R( \{\wb_{j}^{r-1}\}, \Gb)$ in fact lies in the range space of the channel vectors $\{\gb_{kj}\}_j$. Specifically, one can have
\begin{align}\label{transformed gradient form}
	\nabla_{\wb_k} R( \{\wb_{j}\}, \Gb) = \sum_{j=1}^{K}a_{kj}\gb_{kj},
\end{align}
\!\!
where 

\vspace{-0.3cm}
{\small \begin{subequations}\label{eqn: grad coeff}
		\begin{align}
			a_{kk}&= \alpha_{k}\bigg(\sum\limits_{l=1}^K|\gb_{lk}^\Hf\wb_l|^2 + \sigma_k^2\bigg)^{-1}\gb_{kk}^\Hf\wb_k,\\
			a_{kj}&=\frac{-\alpha_{j}|\gb_{jj}^\Hf\wb_j|^2\cdot(\gb_{kj}^\Hf\wb_k)}{\left(\sum_{l=1}^K|\gb_{lj}^\Hf\wb_l|^2 + \sigma_j^2\right) \left(\sum_{l\neq j}|\gb_{lj}^\Hf\wb_l|^2 + \sigma_j^2\right)},j\neq k.
		\end{align}
\end{subequations}}\vspace{-0.3cm}

\noindent {In view of \eqref{transformed gradient form}, it is sufficient to build a learning network that simply learns the $K$ coefficients $\{a_{kj}\}_j$ rather than learning directly the gradient vector $\nabla_{\wb_k} R$.} 

As shown in Fig. \ref{fig:pgddnn}, we construct a beamforming learning network termed ``RNN-PGP" based on the above ideas. 
The learning network has an {RNN} structure with $T$ iterations, each of which mimics the iterative PGP update \eqref{Alg3}-\eqref{eqn: projection 2} for solving problem \eqref{eqn: simplified sum-rate max}.  Specifically, in each iteration $r$, it contains $K$ identical function blocks that produce the beamforming solutions  $\{\wb_k^{r+1}\}$ in parallel.
Here, we illustrate the {detailed} operations of the learning network. 
%
%

{\bf Gradient prediction:} In each iteration, a central preprocessor first gathers the channel information $\{\gb_{kj}\}$, the noise variances $\{\sigma_k^2\}$ and the  beamforming vectors $\{\wb^r_k\}$ obtained in the previous iteration, and then calculates for each $\text{BS}_k$
\begin{align}
	I_j^r &= \sum_{l \neq j}|\gb_{lj}^\Hf\wb_l^r|^2 + \sigma_j^2,~j\in\Kc, \\
	D_j^r&=\alpha_j|\gb_{jj}^\Hf\wb_j^r|^2, ~j\in\Kc,
\end{align} and 
$\{\gb_{kj}^\Hf\wb_k\}_{j\in \Kc}$. The terms $D_j^r$ and $I_j^r$ stand for the information signal power and the suffered interference plus noise power of each $\text{UE}_j$, respectively; while  $\gb_{kj}^\Hf\wb_k$ is the out-going interference of $\text{BS}_k$ to $\text{UE}_j$.
The computed $\{D_j^r,I_j^r,\gb_{kj}^\Hf\wb_k\}_{j\in \Kc}$ is used as the input of a multilayer perception (MLP) block {associated with each $\text{BS}_k$},  which is trained to predict the complex coefficients $\{a_{kj}^r\}_j$ in \eqref{transformed gradient form} and the step size $s_k^r$. Note that the parallel MLPs for all $K$ BSs have an identical network structure with common parameters $\thetab\in \mathbb{R}^{M}$,  where $M$ is the model size. 

The predicted $\{a_{kj}^r\}_j$ are used to construct the gradient vector through the function block $\nabla$
\begin{equation}
\nabla_{\wb_k} R^{r} = \nabla\left(\{a_{kj}^r\}_k, \{\gb_{kj}\}_j\right) = \sum_{j=1}^{K}a_{kj}^r\gb_{kj}.
\end{equation}

{\bf Gradient ascent:}  With $\nabla_{\wb_k} R^{r} $ and $s_k^r$,  the gradient ascent update is performed as
$$\tilde{\wb}_k^{r+1} = {\wb}_k^r + s_k^r \nabla_{\wb_k} R^{r}.
$$ {This step is shown in the middle of the enlarged rectangle in Fig.  \ref{fig:pgddnn}, where} $\otimes$ and $\oplus$ represent the multiplication and addition operations.

{\bf Projection:}  Following \eqref{eqn: projection 2}, the function block $\mathcal{P}$ projects the beamforming vector $\tilde{\wb}_k^{r+1}$ onto the feasible set by computing
\begin{equation}
\hat{\wb}_{k}^{r+1} = \mathcal{P}(\tilde{\wb}_k^{r+1}) = \frac{\tilde{\wb}_{k}^{r+1}}{\max\{\|\tilde{\wb}_{k}^{r+1}\|/\sqrt{P_k}.1\}}.
\end{equation}

{\bf Phase rotation:} It is worth noting that problem \eqref{eqn: simplified sum-rate max} does not have a unique solution. In fact, if $\{\wb_k\}$ is an optimal solution of \eqref{eqn: simplified sum-rate max}, then any phase rotated solution $\{\wb_k e^{i \psi_k}\}$  is also an optimal solution, where $i=\sqrt{-1}$. In order to make sure that the RNN learns a one-to-one mapping, we rotate the phases of 
$\{\hat \wb_k^{r+1}\}$ so that each rotated $\hat \wb_k^{r+1}$, denoted by $\wb_k^{r+1}$, {aligns with $\gb_{kk}$ (i.e., $\gb_{kk}^\Hf{\wb}_k^{r+1}$ is real)}. In particular, we perform via the function block $\angle_{\mathcal{F}}$
\begin{align}
	\wb_{k}^{r+1} &= \angle_{\mathcal{F}}(\hat{\wb}_{k}^{r+1}) \notag \\ &= \hat{\wb}_k^{r+1}\exp{\bigg({-i\tan^{-1}\bigg(\frac{\Im(\gb_{kk}^\Hf\hat{\wb}_k^{r+1})}{\Re(\gb_{kk}^\Hf\hat{\wb}_k^{r+1})}\bigg) }}\bigg).
\end{align}
\begin{figure} [t]
	\centering
	\includegraphics[width=0.4\textwidth]{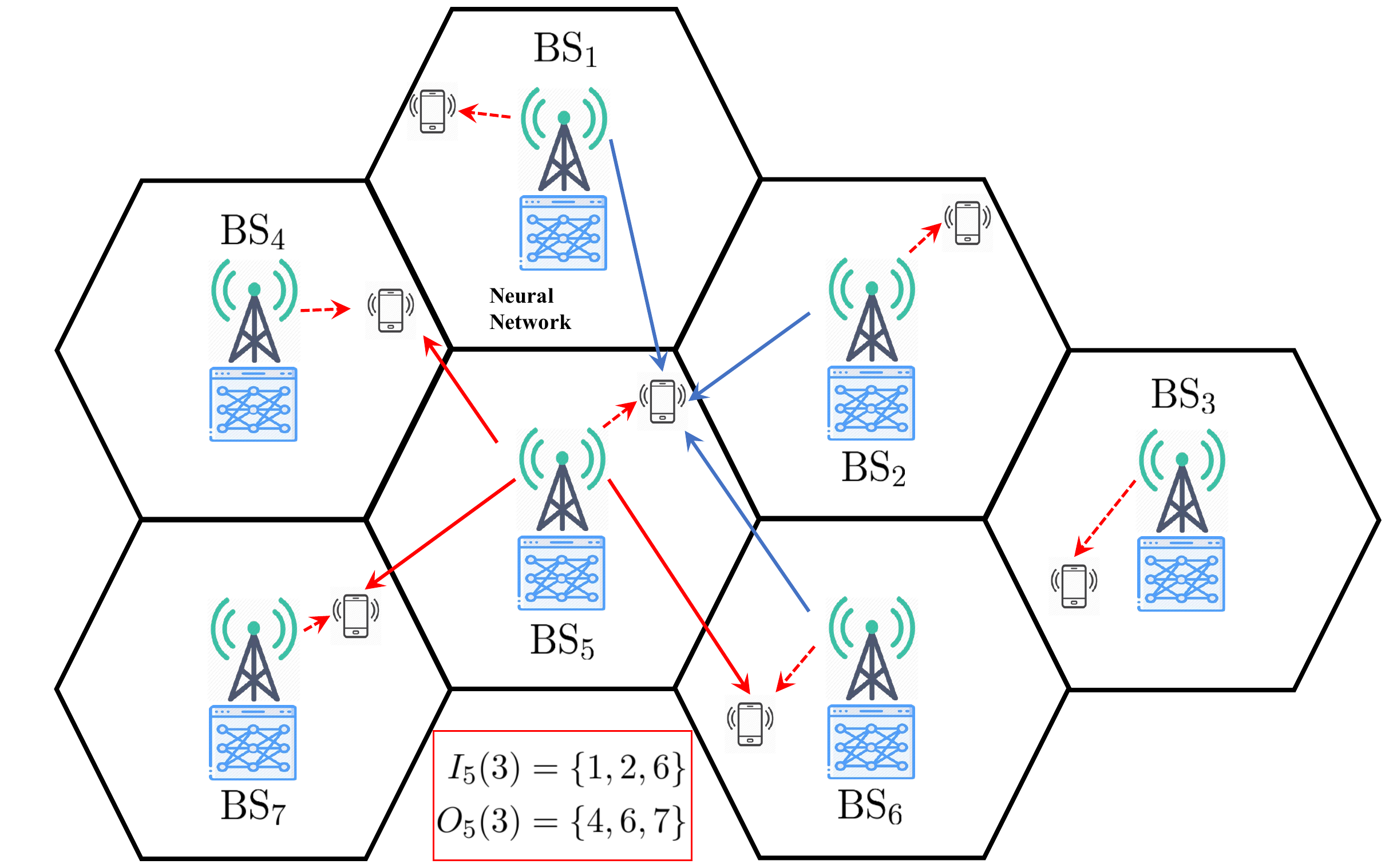}
	\vspace{-0.0cm}
	\caption{\footnotesize Illustration of the neighboring “interferering BSs” and “interfered UEs” of ${\rm BS}_5$. The solid red arrows represent the interference ${\rm BS}_5$ gives to its neighboring “interfered UEs” and the solid blue arrows represent the interference ${\rm UE}_5$ received from neighboring “interferering BSs”.
	}
	\label{fig: system model with limited neighbor}
	\vspace{-0.4cm}
\end{figure}
\begin{figure*} [t]
	\centering
	\includegraphics[width=0.9\textwidth]{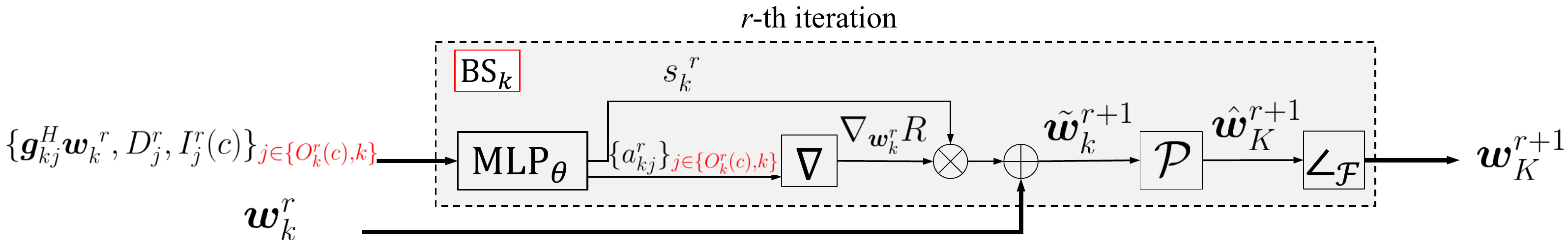}
	\vspace{-0.0cm}
	\caption{\footnotesize Diagram of the revised RNN-PGP network for WSRM problem \eqref{eqn: simplified sum-rate max} where the MLP input and output are based on the subsets $ \mathcal{I}_k^r(c)$ and $ \mathcal{O}_k^r(c) $. Only the block for ${\rm BS}_k$ in the $r$-th iteration is shown in the figure, where $D_j^r = \alpha_j|\gb_{jj}^\Hf\wb_j|^2$ and $I_j^r(c) = \sum_{l \in \mathcal{I}_k^r(c) }|\gb_{lj}^\Hf\wb_l|^2 + \sigma_j^2$.  $\nabla$, $\mathcal{P}$ and $\angle_{\mathcal{F}}$ indicate the gradient prediction, projection and phase rotation respectively.}
	\label{fig:pgddnn neighbor}
	\vspace{-0.4cm}
\end{figure*}

\vspace{-0.3cm}
\subsection{Generalization with {the} Number of BSs}\label{sec: scalable}

As seen in the previous subsection, the proposed RNN-PGP network would have good generalization capability with respect to the number of transmit antennas $N_t$ since both the MLPs and operations in each iteration do not depend on it. {To further make the MLP easy to generalize to settings with different number of BSs $K$}, 
we need the MLP {to have} its input $\{D_j^r, I_j^r, \gb_{kj}^\Hf\wb_k^r\}_{j\in \Kc}$ and output $\{a_{kj}^r\}_{j\in \Kc}$ to be independent of the BS number $K$. 

As inspired by \cite{nasir2019multi}, for each ${\rm BS}_k$
we define two neighbor subsets. The first is the set of BSs that cause a relatively strong interference to ${\rm BS}_k$, i.e.,
\begin{align}\label{eqn: interfering}
	\mathcal{I}_k^r \coloneqq \left\{j \in \Kc, j\neq k \big|\ |\gb_{jk}^\Hf\wb_j^r|^2 > \eta\sigma_{k}^2\right\},
\end{align} where $\eta$ a preset threshold. We order 
$\ |\gb_{jk}^\Hf\wb_j^r|^2/\sigma_{k}^2$, $j\in \mathcal{I}_k^r$ in a decreasing fashion, and further select the first $c$ indices in  $ \mathcal{I}_k^r$. The selected subset is denoted by $ \mathcal{I}_k^r(c) \subseteq \mathcal{I}_k^r$, which indicates the first $c$ most ``interferering”  BSs that have strong interference on ${\rm BS}_k$.

Analogously,  for each ${\rm BS}_k$ we define in the following the set of  UEs whom ${\rm BS}_k$ causes strong interference to 
\begin{align}\label{eqn: interfered}
	\mathcal{O}_k^r \coloneqq \left\{j \in \Kc, j\neq k \big|\ |\gb_{kj}^\Hf\wb_k^r|^2 > \eta\sigma_{j}^2 \right\},
\end{align}
and further select a subset $ \mathcal{O}_k^r(c) \subseteq \mathcal{O}_k^r$ which corresponds to the $c$ most ``interfered” UEs by ${\rm BS}_k$. An example is shown in Fig. \ref{fig: system model with limited neighbor}, where $c=3$, and the ``interferering” neighbors of $\text{BS}_5$ are $\{\text{BS}_1, \text{BS}_2, \text{BS}_6\}$ and “interfered” UEs of it are $\{\text{UE}_4, \text{UE}_6, \text{UE}_7\}$.

Then, we consider only BSs in $ \mathcal{I}_k^r(c)$ and UEs in $ \mathcal{O}_k^r(c) $ for computing the input and output of the MLP associated with $\text{BS}_k$. 
Specifically, instead of $\{D_j^r,I_j^r,\gb_{kj}^\Hf\wb_k\}_{j\in \Kc}$, we let the input of the MLP associated with ${\rm BS}_k$ be $\{D_j^r, I_j^r(c),\gb_{kj}^\Hf\wb_k^r\}_{j\in\{\mathcal{O}^r_k(c), k\}}$, where 
\begin{align}\label{eqn: neighbor interference}
	I_j^r(c) = \sum_{l \in \mathcal{I}_j^r(c)}|\gb_{lj}^\Hf\wb_l^r|^2 + \sigma_j^2.
\end{align}
In addition to the step size $s_k^r$, the output of the MLP associated with ${\rm BS}_k$ 
is changed to $\{a_{kj}^r\}_{j\in\{\mathcal{O}^r_k(c), k\}}$, {which now is dependent on the set $\mathcal{O}^r_k(c)$ only}. Therefore, the input and output of the MLP are independent of the whole network size $K$. The diagram of the revised RNN-PGP network based on $\mathcal{I}_k^r(c)$ and $ \mathcal{O}_k^r(c) $
is illustrated in Fig. \ref{fig:pgddnn neighbor}, where only the block associated with ${\rm BS}_k$ at iteration $r$ is plotted. 

%
%
%
%
%
\vspace{-0.3cm}
\subsection{Hybrid Training Strategy}\label{subsec: hybrid training}
Suppose that a training data set of size $L$ is given, which contains the (dimension reduced) CSI $\{\Gb^{(\ell)}\}$, WSR coefficients $\{\alpha_k^{(\ell)}\}_{k \in \Kc}$, and the beamforming solutions $\{\wb_k^{(\ell)}\}_{k\in \Kc}$ obtained by an existing algorithm, for $\ell=1,\ldots, L$. 
Suppose that the RNN-PGP has a total number of $T$ iterations (see Fig. \ref{fig:pgddnn}), and denote $\{\wb_k^{(\ell), r}\}_{k\in \Kc}$ as the output of RNN-PGP in the $r$th iteration for the $\ell$th data sample.  
The RNN-PGP network is trained by a two-stage approach. The first stage is based on supervised learning, using the following loss function 
\begin{equation}\label{eqn: loss penalty supervised}
\begin{aligned}
\text{L}^{\text{S}}(\thetab) = \frac{1}{2LK}&\sum_{\ell=1}^{L}\sum_{k=1}^{K}\alpha_k^{(\ell)}\Bigg(\gamma\lVert \wb_{k}^{(\ell)} - \wb_k^{(\ell), T}\rVert^2\\ 
&+ (1-\gamma)\sum_{r=1}^{T-1}\lVert \wb_{k}^{(\ell)} - \wb_k^{(\ell), r}\rVert^2\Bigg) ,
\end{aligned}\end{equation}
where  
$\gamma$ is the penalty parameter. Note that the second term in the right hand side of \eqref{eqn: loss penalty supervised} encourages the output of earlier iterations of RNN-PGP to be close to $\{\wb_k^{(\ell)}\}$, which may help speed up the convergence of RNN-PGP.

By treating the supervised training in the first stage as a pre-training, we can further refine the network in an unsupervised fashion in the second stage. Specifically, we can directly train the network so that the WSR function of \eqref{eqn: miso sum-rate maximization} is maximized; for example, we consider the following loss function 
\begin{equation}\label{eqn: loss unsupervised}
\text{L}^{\text{U}}(\thetab) = -\frac{1}{KL}\sum_{\ell =1}^{L}R(\{\wb_{k}^{(\ell), T}\},\Gb^{(\ell)}).
\end{equation}
As will be shown in Section \ref{sec:simulation}, the two-stage training approach can outperform those that solely use supervised training or unsupervised training \cite{xia2019deep}.  

\vspace{-0.3cm}
\section{Extension to Cooperative Multicell Beamforming Problem}\label{sec:other applications} 
The MISO interference channel considered in Section \ref{sec: system model} and Section \ref{sec: PGP Inspired RNN} treats the interference from adjacent cells as noise, 
resulting in a fundamental limitation on the performance especially for terminals close to cell 
edges \cite{karakayali2006network}. 
In recent years, BS coordination has been analyzed as a means of handling inter-cell interference, in which one UE is served by multiple BSs \cite{zhang2004cochannel}. 

In this section, we extend the RNN-PGP network to the cooperative multicell scenario.
As shown in Fig. \ref{fig:fig2}, the cooperative multicell communication scenario considered herein consists of $K_r$ single-antenna receivers served by $K_t$ BSs equipped with $N_t$ antennas each. The $j$th transmitter and $k$th receiver are denoted $\text{BS}_j$ and $\text{UE}_k$, respectively; the channel between them is denoted by $\hb_{jk}$ for $j\in \Kc_t\coloneqq \{1,\ldots,K_t\}$ and $k\in \Kc_r\coloneqq \{1,\ldots,K_r\}$.
Let $\vb_{jk}$ be the beamforming vector used by $\text{BS}_j$ for serving $\text{UE}_k$. 
The SINR at $\text{UE}_k$ is given by
\begin{equation}\label{eqn: SINR}
\SINR_k = \frac{\left|\sum_{j=1}^{K_t}\hb_{jk}^\Hf\vb_{jk}\right|^2}{\sum_{l\neq k}^{K_r}\left|\sum_{j=1}^{K_t}\hb_{jk}^\Hf\vb_{jl}\right|^2+\sigma_k^2}.
\end{equation}
The WSRM problem is formulated as
\begin{subequations}\label{eqn: coop sum-rate maximization}
	\begin{align}
		\max_{ \substack{\{\vb_{jk}\}_{j \in \Kc_t,k\in\Kc_r}}}~  & \sum_{k=1}^{K_r} \alpha_k\log_2\left(1 + \SINR_k \right) \\
		{\rm s.t.}~ & \sum_{k=1}^{K_r}\lVert \vb_{jk}\rVert^2 \leq P_j, j\in \Kc_t, \label{coop power constraint}
	\end{align}
\end{subequations}
where \eqref{coop power constraint} represents the total power constraint of each $\text{BS}_j$.

Analogous to Section \ref{sec: PGP Inspired RNN}, we employ  \cite[Theorem 2]{bjornson2010cooperative} to transform 
problem \eqref{eqn: coop sum-rate maximization} into a dimension-reduced problem.
\begin{theo}
	\cite[Theorem 2]{bjornson2010cooperative}\label{coop beamform linear structure}
	For each rate tuple on the Pareto boundary for problem \eqref{eqn: coop sum-rate maximization}, it holds that 
	beamformers $\{\vb_{jk}\}$ that achieve the Pareto boundary fulfill 
	\begin{equation}\label{theorem2}
	\vb_{jk} \in \text{span}\left(\{\hb_{jk}\}\bigcup_{l\neq k}\left\{\Pi_{\hb_{jl}}^{\perp} \hb_{jk}\right\}\right), \forall j,k,
	\end{equation}
	where $\Pi_{\hb_{jl}}^{\perp} \coloneqq \Ib_{N_t} - \hb_{jl}\hb_{jl}^\Hf/\|\hb_{jl}\|^2$ is the orthogonal projection onto the orthogonal complement of $\hb_{jl}$.
\end{theo}
Based on Theorem \ref{coop beamform linear structure},  the optimal beamforming solution $\vb_{jk}, j\in \Kc_t,k\in\Kc_r$ of problem \eqref{eqn: coop sum-rate maximization} can be expressed as 
\begin{align}
	\vb_{jk} &= \xi_{jk}^{k}\hb_{jk} + \sum_{l\neq k}^{K_r}\xi_{jk}^{l}\hb_{jk}^{l\perp} \notag \\
	& = \Hb_{jk} \xib_{jk},
\end{align}
where $\hb_{jk}^{l\perp} \coloneqq \Pi_{\hb_{jl}}^{\perp}\hb_{jk}$, $\xib_{jk}=\begin{bmatrix} \xi_{jk}^1, \ldots, \xi_{jk}^{K_r} \end{bmatrix}\in \Cbb^{K_r}$ and 
$\Hb_{jk}=\begin{bmatrix} \hb_{jk}^{1\perp}, \ldots, \hb_{jk}^{k-1\perp},\hb_{jk},\hb_{jk}^{k+1\perp},\ldots, \hb_{jk}^{K_r\perp} \end{bmatrix}  \in \Cbb^{N_t\times K_r}$.
Further consider the eigenvalue decomposition of 
$$\Hb_{jk}^\Hf\Hb_{jk}=\Ub_{jk}\LambdaB_{jk}\Ub_{jk}^\Hf,$$
and define $\wb_{jk} = \LambdaB_{jk}^{1/2}\Ub_{jk}^\Hf\xib_{jk}$, and $\gb_{jk} = (\LambdaB_j)^{-1/2}\Ub_j^{\Hf}\Hb_{jk}^\Hf\hb_{jk} $. We can rewrite \eqref{eqn: coop sum-rate maximization} as

\vspace{-0.3cm}
{\small \begin{align}\label{eqn: simplified sum-rate max coop}
		\max_{ \substack{\{\wb_{jk}\}_{j \in \Kc_t,k\in\Kc_r}}} & \sum_{k=1}^{K_r} \log_2\left(1 + \frac{\left|\sum_{j=1}^{K_t}\gb_{jk}^\Hf\wb_{jk}\right|^2}{\sum_{l\neq k}^{K_r}\left|\sum_{j=1}^{K_t}\gb_{jk}^\Hf\wb_{jl}\right|^2+\sigma_k^2} \right) \notag \\
		{\rm s.t.}~ & \sum_{k=1}^{K_r}\lVert \wb_{jk}\rVert^2 \leq P_j, j\in \Kc_t.
\end{align}}\vspace{-0.3cm}

%


Let us slightly abuse the notation by defining $R(\{\wb_{jk}\},\Gb)$ as the WSR in \eqref{eqn: simplified sum-rate max coop}. The gradient of $R(\{\wb_{jk}\},\Gb)$ with respect to each $\wb_{jk}$ has the following form
\begin{equation}\label{eqn: gradient of coop}
\nabla_{\wb_{jk}} R = \sum_{p=1}^{K_r}\left(a_{jp}^{(k)}\gb_{jp} + b_{jp}^{(k)}\gb^*_{jp}\right),
\end{equation}
where $a_{jp}^{(k)} = c_{jp}^{(k)}\gb_{jp}^\Hf\wb_{jk}$ and $b_{jp}^{(k)} = c_{jp}^{(k)} \sum_{q\neq j}^{K_t}\gb_{qp}^\Tf\wb_{qk}^*$, in which $c_{jk}^{(k)}=\sum_{l =1 }^{K_r}\left|\sum_{q=1}^{K_t}\gb_{qk}^\Hf\wb_{ql}\right|^2+\sigma_k^2$ and 
\begin{align*}\small
	c_{jp}^{(k)} = -\frac{\left|\sum\limits_{q=1}^{K_t}\gb_{qp}^\Hf\wb_{qp}\right|^2}{\left(\sum\limits_{l \neq p }^{K_r}\left|\sum\limits_{q=1}^{K_t}\gb_{qp}^\Hf\wb_{ql}\right|^2+\sigma_p^2\right)\left(\sum\limits_{l=1 }^{K_r}\left|\sum\limits_{q=1}^{K_t}\gb_{qp}^\Hf\wb_{ql}\right|^2+\sigma_p^2\right)},
\end{align*}
for all  $p\neq k$.
Therefore, similar to Fig. \ref{fig:pgddnn}, we can build a learning network for problem \eqref{eqn: simplified sum-rate max coop} by unfolding the PGP method and learning the complex coefficients $\{a_{jp}^{(k)},b_{jp}^{(k)}\}$ in \eqref{eqn: gradient of coop}.

\begin{figure*} [t]
	\centering
	\includegraphics[width=0.7\textwidth]{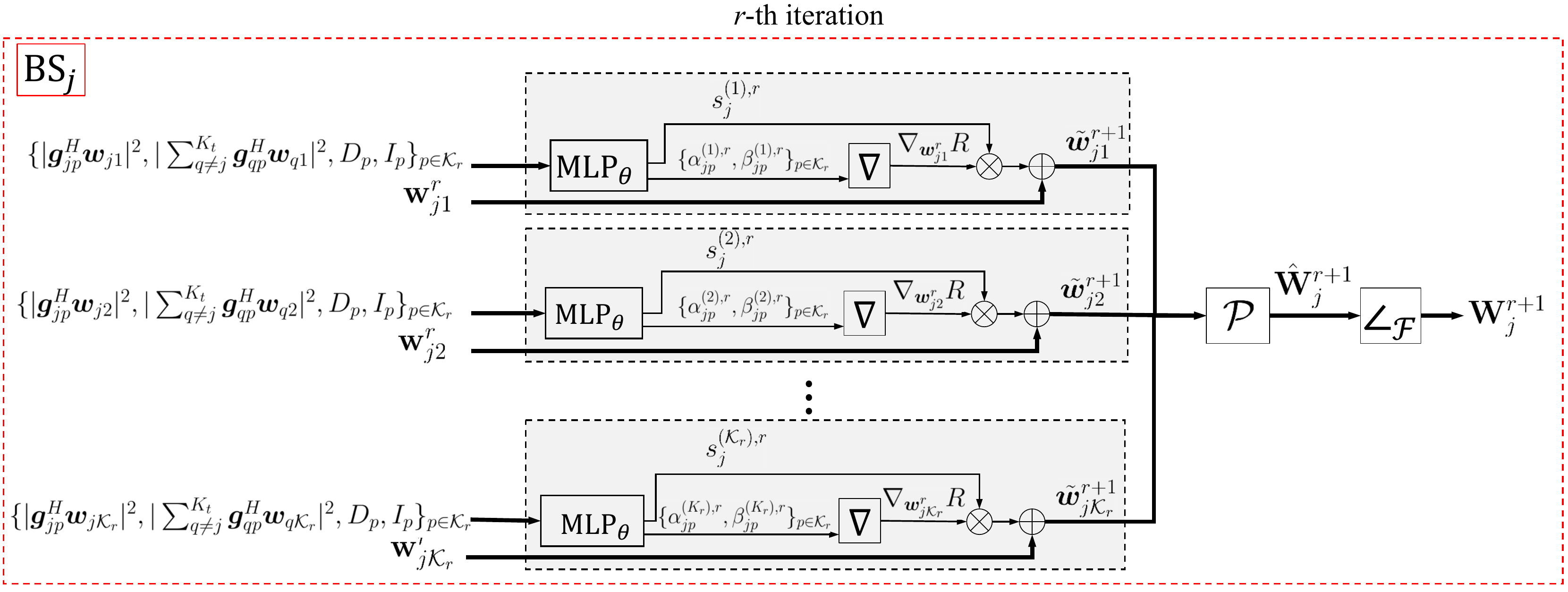}
	\vspace{-0.0cm}
	\caption{\footnotesize Diagram of the proposed RNN-PGP network for WSRM problem \eqref{eqn: simplified sum-rate max coop} of the cooperative multicell scenario. The subscript $(\cdot)^{(k),r}_{jp}$ refers to the parameters in the $j$-th BS for the $k$-th user in the $r$-th iteration. $
		\hat{\Wb}_{j} = [\hat{\wb}_{j1},\ldots,\hat{\wb}_{jK_r}]$, $\Wb_{j} = [\wb_{j1},\ldots,\wb_{jK_r}]$. The subscript $(\cdot)^r_k$ refers to the parameter for the $k$-th BS in the $r$-th iteration. $\nabla$, $\mathcal{P}$ and $\angle_{\mathcal{F}}$ indicate the gradient, projections and phase rotation respectively.}
	\label{fig:coop-rnn-network}	\vspace{-0.4cm}
\end{figure*}

In Fig. \ref{fig:coop-rnn-network}, we present the block diagram of the RNN-PGP network for problem \eqref{eqn: simplified sum-rate max coop}, where we only plot the function blocks associated with ${\rm BS}_j$ at the $r$th iteration.
The RNN-PGP network for the cooperative multicell scenario has a similar structure as that in Fig. \ref{fig:pgddnn}. 
All the MLPs also share the same structure and parameters. 
The input of the MLP of ${\rm BS}_j$ for ${\rm UE}_k$
is $\{|\gb_{jp}^\Hf\wb_{jk}^r|^2, |\sum_{q\neq j}^{K_t}\gb_{qp}^\Hf\wb_{qk}^r|^2, D_p^r, I_p^r\}_{p\in\Kc_r}$, where
\begin{equation}
\begin{aligned}
D_p^r &= \bigg|\sum_{q=1}^{K_t}\gb_{qp}^\Hf\wb_{qp}^r\bigg|^2,~
I_p^r &=\sum_{l\neq p}^{K_r} \bigg|\sum_{q=1}^{K_t}\gb_{qp}^\Hf\wb_{ql}^r\bigg|^2 + \sigma_p^2.
\end{aligned}
\end{equation}
The output of the MLP of ${\rm BS}_j$ for ${\rm UE}_k$ in the $r$th iteration is 
$\{a_{jp}^{(k),r}, b_{jp}^{(k),r}\}_{p\in\Kc_r}$ and  the step size $s_{jk}^{(k), r}$.

Then, the block $\nabla$ constructs the gradient vector according to \eqref{eqn: gradient of coop}, i.e.,
\begin{align}
	\nabla_{\wb_{jk}} R^r &= \nabla(\{a_{jp}^{(k),r}, b_{jp}^{(k),r}\}_p, \{\gb_{jp}\}_p)\\
	& = \sum_{p=1}^{K_r}\left(a_{jp}^{(k),r}\gb_{jp} + b_{jp}^{(k),r}\gb^*_{jp}\right),~k\in \Kc_r,
\end{align}
followed by gradient ascent update $\tilde{\wb}_{jk}^{r+1} = {\wb}_{jk}^{r} + s_{jk}^r \nabla_{\wb_{jk}} R^r$, $k\in \Kc_r$.

All the beamforming vectors $\{\tilde{\wb}_{jk}^{r+1}\}_k$ due to ${\rm BS}_j$ will be collected and {used to} perform projection onto the feasible set of  problem \eqref{eqn: simplified sum-rate max coop}, which yields
\begin{equation}
\hat{\wb}_{jk}^{r+1} = \mathcal{P}(\tilde{\wb}_{jk}^{r+1}) = \frac{\tilde{\wb}_{jk}^{r+1}}{\max\{\sqrt{\sum_{k=1}^{K_r}\|\tilde{\wb}_{jk}^{r+1}\|^2/P_j},1\}},
\end{equation} 
for all $k\in \Kc_r$.
Lastly, the function block $\angle_{\mathcal{F}}$ rotates the phases of $\{\hat{\wb}_{jk}^{r+1}\}_k$ by
\begin{align}
	\wb_{jk}^{r+1} &= \angle_{\mathcal{F}} (\hat{\wb}_{jk}^{r+1}) \notag \\
	&= \hat{\wb}_{jk}^{r+1}\exp{\bigg({-i\tan^{-1}\bigg(\frac{\Im(\gb_{jk}^\Hf\hat{\wb}_{jk}^{r+1})}{\Re(\gb_{jk}^\Hf\hat{\wb}_{jk}^{r+1})}\bigg)}\bigg)}.
\end{align}

To train the RNN-PGP network in Fig. \ref{fig:coop-rnn-network}, we adopt the same hybrid strategy in Section \ref{subsec: hybrid training}. The supervised training loss and the unsupervised training loss are respectively given by
\begin{align}\label{eqn: loss penalty supervised coop}
	\text{L}^{\text{S}}(\thetab) = \frac{1}{2LK_tK_r}&\sum_{\ell=1}^{L}\sum_{j=1}^{K_t}\sum_{k=1}^{K_r}\Big(\gamma\lVert \wb_{jk}^{(\ell)} - \wb_{jk}^{(\ell), T}\rVert^2 \notag\\
	& + (1-\gamma)\sum_{r=1}^{T-1}\lVert \wb_{jk}^{(\ell)} - \wb_{jk}^{(\ell),r}\rVert^2\Big).
\end{align}
and 
\begin{equation}\label{eqn: loss unsupervised coop}
\text{L}^{\text{U}}(\thetab)= -\frac{1}{LK_r}\sum_{l=1}^{L}R(\{\wb_{jk}^{(\ell), T}\},\Gb^{(\ell)}).
\end{equation}

We remark that, similar to that in Fig. \ref{fig:pgddnn}, the RNN-PGP network in Fig. \ref{fig:coop-rnn-network} for the coopeartive multicell scenario inherently has a good generalization with respect to the number of transmit antennas $N_t$. Besides, since the input size and output size of the MLPs in Fig. \ref{fig:coop-rnn-network} do not depend on the number of cooperative BSs $K_t$, the RNN-PGP network also has good generalization capability with respect to $K_t$. These will be examined in the simulation results in Section \ref{sec:simulation}.


\vspace{-0.2cm}
\section{Simulation Results}\label{sec:simulation}
In this section, we present numerical results of the proposed RNN-PGP networks in Fig. \ref{fig:pgddnn} and Fig. \ref{fig:coop-rnn-network}. Both the synthetic channel models and the ray-tracing based DeepMIMO dataset \cite{alkhateeb2019deepmimo} are considered to examine the performance of the proposed beamforming learning networks.

\vspace{-0.3cm}
\subsection{Simulation Setup}
We first consider the MISO interference channel model in Section \ref{sec: system model} and test the performance of the  proposed RNN-PGP network in Fig. \ref{fig:pgddnn} under synthetic Rayleigh channel data.  
Like Fig. \ref{fig: system model with limited neighbor}, we assume that each ${\rm BS}_k$ is located at the center of cell $k$ and ${\rm BS}_k$ is located randomly according to a uniform distribution within the cell. 
The half BS-to-BS distance is denoted as $d$ and it is chosen between $100$ and $1000$ meters. 
We set $P_{max}$, i.e., the maximum transmit power level of ${\rm BS}_k$, to be $38$ dBm over a $10$ MHz frequency band. The path loss between the UE and its associated BS is set as $128.1 + 37.6\log_{10}(s)$ (dB) where $s$ (km) is the distance between the UE and BS.  The channel coefficients of $\{\hb_{kj}\}$ are generated following the independent and identically distributed complex Gaussian distribution with mean equal to the pathloss and variance equal to one.
The noise power spectral density of all UEs are set the same and equal to $\sigma^2 = -174$ dBm/Hz.
%
A total of $5000$ training samples ($L=5000$) are generated which include CSI $\{\hb_{kj}^{(\ell)}\}$, $\ell=1,\ldots,L$, rate weights $\{\alpha_k^{(\ell)}\}$, $\ell=1,\ldots,L$, and 
beamforming solutions $\{\wb_k^{(\ell)}\}$, $\ell=1,\ldots,L$, obtained either by the PGP method or the POA algorithm.
Another $1000$ testing samples are also generated in the same way.

To train the RNN-PGP network in Fig. \ref{fig:pgddnn}, {we set
	the parameter $\eta$ in \eqref{eqn: interfering} and \eqref{eqn: interfered} to be $5$, and the parameter $\gamma$ in \eqref{eqn: loss penalty supervised} and \eqref{eqn: loss penalty supervised coop} to be 0.95.}
The function $\tanh$ is used as the activation function in the MLP.
The iteration number of RNN-PGP is set to $T=20$, 
and the Adam optimizer is used for training the RNN-PGP network. The simulation environment is based on Python 3.6.9 with TensorFlow 1.14.0 on a desktop computer with Intel i7-9800X CPU Core, one NVIDIA RTX 2080Ti GPU, and 64GB of RAM. The GPU is used during the training stage but not  used in the test stage. 

In the presented results, we benchmark the proposed RNN-PGP network with the existing beamforming algorithms such as the PGP method, WMMSE algorithm and the POA algorithm, as well as black-box based DNN approach. Specifically, they include 
\begin{itemize}
	\item {\bf PGP, WMMSE and POA:} Performance results obtained by applying the three methods to the dimension reduced problem \eqref{eqn: simplified sum-rate max}, respectively.
	
	\item {\bf  RNN-PGP (PGP) and RNN-PGP (POA):} The  RNN-PGP networks in Fig. \ref{fig:pgddnn} trained by the hybrid strategy, and the beamforming solutions used for supervised training are obtained by the PGP method and the POA method, respectively. 
	
	\item {\bf  DNN (PGP) and DNN (POA):} The black-box DNNs, which have $5$ layers with the concatenated CSI as the input and the concatenated beamforming solutions as the output, 
	are trained end-to-end by the hybrid training strategy, and the beamforming solutions used for supervised training are obtained by the PGP method and the POA method, respectively.

	
	
	
	\item {\bf  RNN-PGP (Unsuper):} The RNN-PGP network in Fig. \ref{fig:pgddnn} trained solely by the unsupervised cost.
	
	\item  {\bf RNN-PGP (POA, Stepsize):} The MLPs in the RNN-PGP only predicts the step size $s_k^r$, and the gradient vector $\nabla_{\wb_k} R^{r} $ is computed explicitly by \eqref{transformed gradient form} and  \eqref{eqn: grad coeff}. The network is trained by the hybrid strategy and the beamforming solutions obtained by the POA method is used during the supervised training.
\end{itemize}
For the presented test results, if not mentioned specifically, the beamforming solutions of RNN-PGP are obtained by unfolding the network for $T=20$ iterations. 
Except for the (weighted) sum rate, we also show the ``accuracy" (\%) which is the ratio of the (weighted) sum rate achieved by the RNN-PGP and that achieved by the WMMSE solution.
%

%

\vspace{-0.3cm}
\subsection{Sum-rate Performance}
\begin{figure}[t]
	\centering
	\subfigure[Sum-rate versus iteration.]{
		\begin{minipage}{0.4\textwidth}
			\centering
			\includegraphics[width=\textwidth]{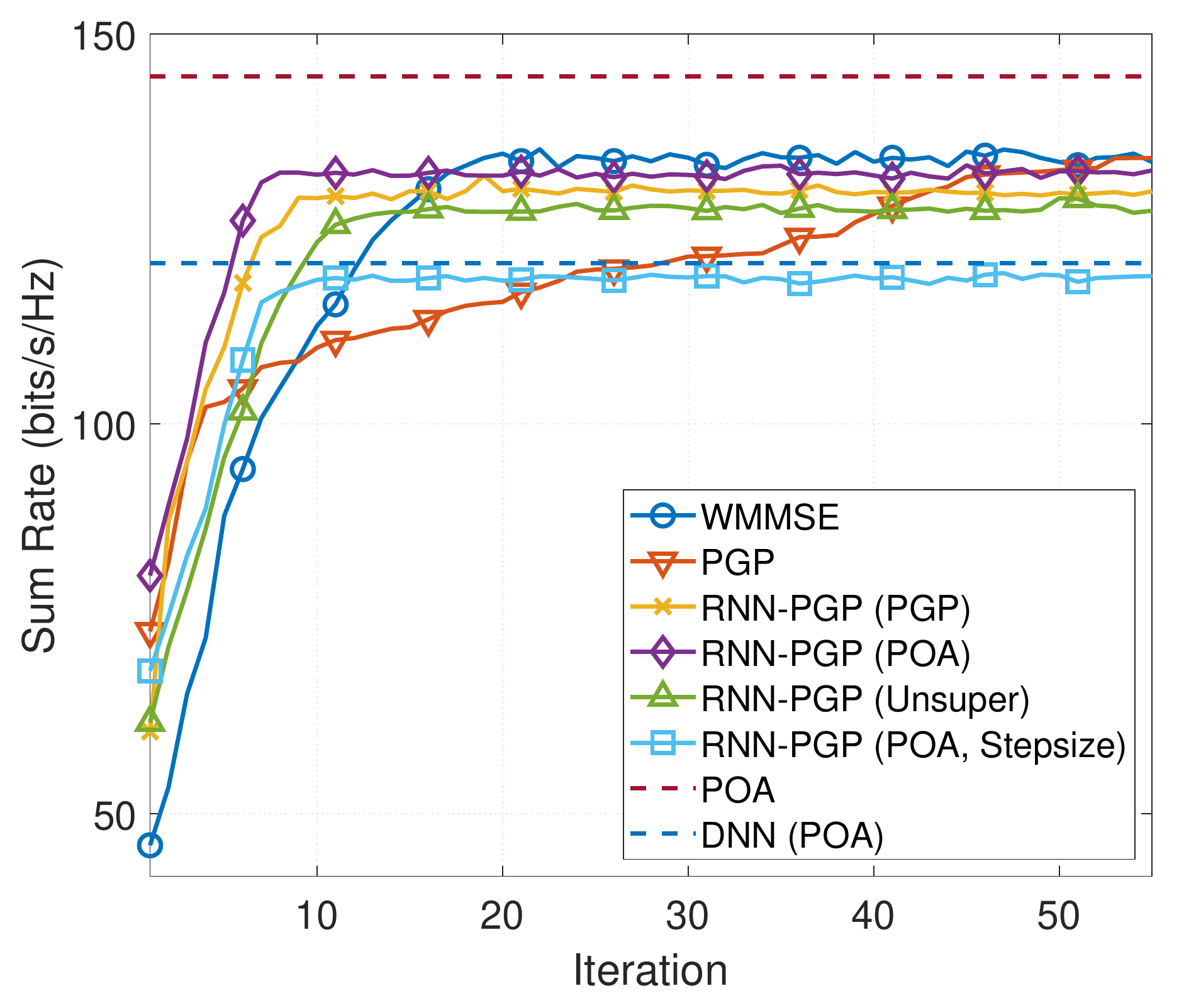}
			\label{fig: Sum-rate versus iteration}	
	\end{minipage}}
	\subfigure[Sum-rate versus runtime for $T=55$ iterations.]{
		\begin{minipage}{0.4\textwidth}
			\centering
			\includegraphics[width=\textwidth]{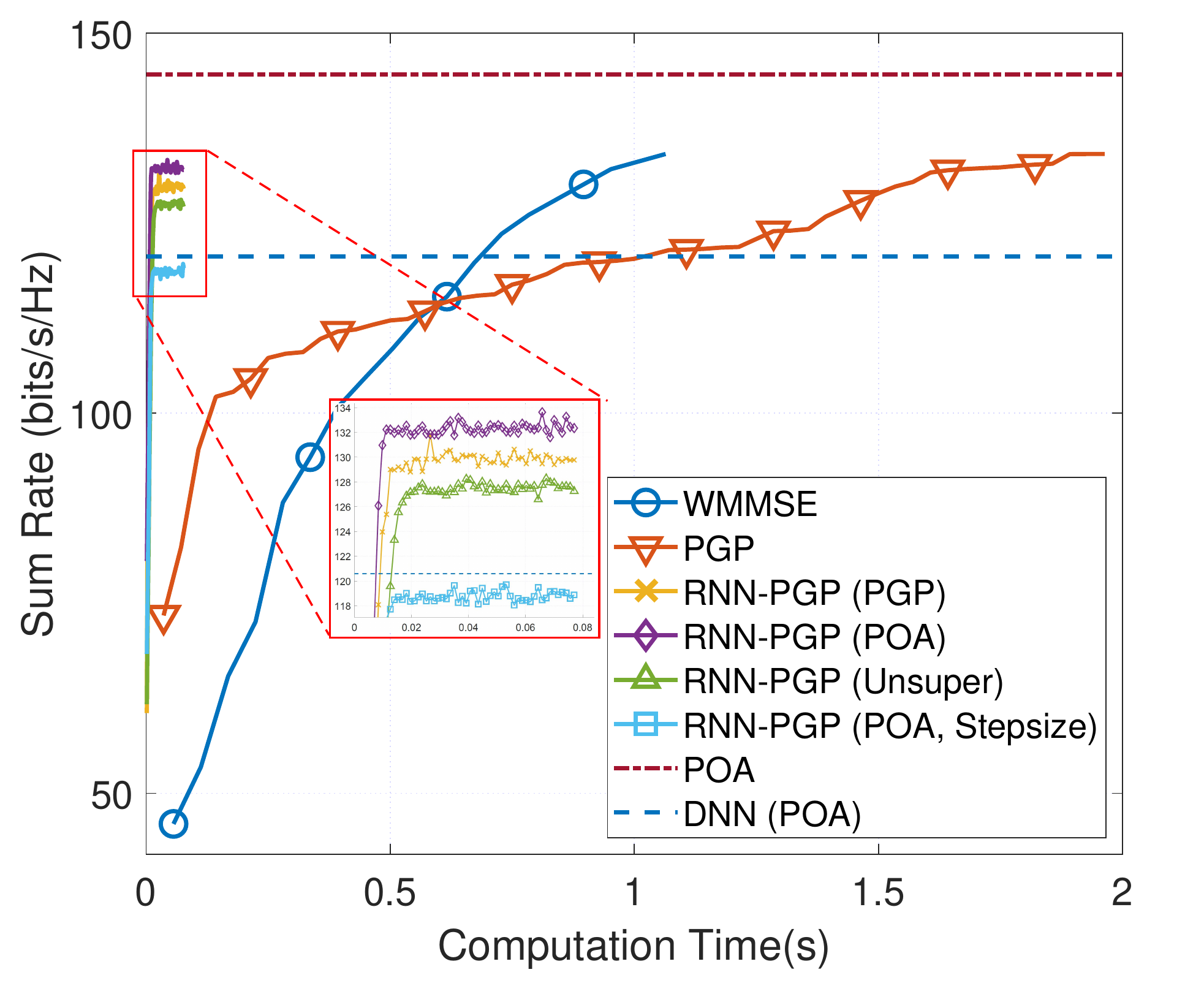}
			\label{fig: Sum-rate versus runtime}	
	\end{minipage}}
	\caption{\footnotesize The sum rates achieved by various schemes versus the iteration number and  runtime.}\vspace{-0.4cm}
\end{figure}

In this subsection, we evaluate the sum-rate performance of different schemes by setting the weights $\alpha_k$ of all users to be $1$. The results are shown in Fig. \ref{fig: Sum-rate versus iteration}, Fig. \ref{fig: Sum-rate versus runtime} and Tables \ref{table: number of antennas}, \ref{table:number of bs}, \ref{table:number of neighbors}, respectively.  

If not mentioned specifically, we consider the MISO interference channel with $K = 19$ and $N_t = 36$. We set the neighbor size $c$ to be 18 ($c=18$). 
For the ``RNN-PGP (PGP)", ``RNN-PGP (POA)" and ``RNN-PGP (Unsuper)", the MLP used is a 5-layers DNN with the numbers of neurons of the input layer and output layer are $4K$ and $2K+1$,  respectively, and the numbers of neurons in the three hidden layers are $125, 100, 85,$ respectively. For the ``DNN (PGP)" and ``DNN (POA)", we use a 5-layers DNN where the number of nodes of the input and output layers are both $2KN_t$, and the numbers of nodes in the hidden layers are $1450, 1250, 1325$, respectively. 
The input of the MLP in the ``RNN-PGP (POA, Step-size)" is the same as that of the ``RNN-PGP (PGP)", but the numbers of neurons of the hidden layer are $120, 75, 25$, respectively, and the number of neurons of the output is reduced to $1$.

%

{\bf Convergence and runtime:} 
The experiment results of the achieved sum rates versus the iteration number are shown in Fig. \ref{fig: Sum-rate versus iteration}, where the RNN-PGPs are unfolded for $55$ iterations. For the POA algorithm and the ``DNN (POA)", we simply plot the lines indicating the sum rates achieved by the two methods. Firstly, one can observe that the POA algorithm provides a sum-rate upper bound, and that the WMMSE algorithm not only converges faster but also yields a slighter higher sum rate than the PGP method.

Secondly, we can see that ``RNN-PGP (POA)" converges faster and yields higher sum rates than ``RNN-PGP (PGP)" as well as  ``RNN-PGP (Unsuper)" and ``RNN-PGP (POA, Stepsize)", which shows the benefits of the hybrid training strategy and prediction of the gradient vector. 
Note from the figure that 
both ``RNN-PGP (POA)"  and ``RNN-PGP (PGP)" can converge well around 10 iterations and achieve comparable sum rate as the WMMSE and PGP algorithms. 
Thirdly, except for ``RNN-PGP (POA, Stepsize)", the RNN-PGP networks can greatly outperform the black-box based ``DNN (POA)".

\renewcommand\arraystretch{1.2}
\begin{table}[t]
	\centering
	\caption{The sum-rate performance of the proposed RNN-PGP for $K = 19$ and $c=18$; the size of hidden layers of the  MLP are 125, 100, and 85 respectively (MLP 125:100:85).}
	\label{table: number of antennas}
	\resizebox{\textwidth}{!}{%
		\begin{tabular}{cccccc}
			\toprule
			\multicolumn{2}{c}{\textbf{Number of antennas}}                                                                  & $N_t =36$                                       & $N_t = 72$                                      &$N_t = 108$                                      & $N_t$ randomly selected from 16-128          \\
			\rowcolor[gray]{.8} 
			\multicolumn{2}{c}{\cellcolor[gray]{.8}\textbf{PGP}}                                            & 134.21                                   & 145.16                                  & 151.19                                   & -                                   \\
			& Trained with $N_t=36$                              & 129.85 (96.75\%)                         & 137.29 (94.58\%)                        & 142.81 (94.46\%)                         & - (93.78\%)                         \\
			\multirow{-2}{*}{\textbf{RNN-PGP (POA)}} & \cellcolor[gray]{.8}Trained with mixed $N_t \in \{18,\ 36,\ 72,\ 108\}$ & \cellcolor[gray]{.8}128.34 (95.63\%) & \cellcolor[gray]{.8}138.47 (95.39\%) & \cellcolor[gray]{.8}143.89 (95.17\%) & \cellcolor[gray]{.8}- (95.27\%)\\
			\bottomrule
		\end{tabular}%
	}
\end{table}
\renewcommand\arraystretch{1.2}
\begin{table}[t]
	\caption{The sum-rate performance of the proposed RNN-PGP for $N_t=64$.}
	\centering
	\subtable[]{        
			\resizebox{\textwidth}{!}
				{
		\begin{tabular}{clcccc}
			\toprule	
			\multicolumn{2}{c}{\textbf{Number of Neighbors}}                                                                                                   & \multicolumn{4}{c}{$c=18\ (\text{MLP}\ 85:73:42)$}                                                                                                                                             \\
			\rowcolor[gray]{.8}
			\multicolumn{2}{c}{\cellcolor[gray]{.8}\textbf{Number of BS-user links}}                                                                           & $K=37$                                  & $K=61$                                  & $K=91$                                & $K=91\ (N_t=128)$                         \\
			&Trained with $K=37$  & 221.88 (96.01\%) &289.21 (93.12\%) &535.33 (91.92\%) &596.57 (91.79\%) \\
			\multirow{-2}{*}{\textbf{RNN-PGP (POA)}} &  \cellcolor[gray]{.8} Trained with $K \in\{37,\ 61,\ 91\}$                          & \cellcolor[gray]{.8}219.85 (95.13\%)                         & \cellcolor[gray]{.8}291.48 (93.85\%)                         & \cellcolor[gray]{.8}540.68 (92.84\%)                         &\cellcolor[gray]{.8} 598.07 (92.02\%)   \\
			\bottomrule                   
	\end{tabular}}
	\label{table:number of bs-b}}
	\\
	\subtable[]{
				\resizebox{\textwidth}{!}{
		\begin{tabular}{clcccc}
			\toprule	
			\multicolumn{2}{c}{\textbf{Number of Neighbors}}                                                                                                  & \multicolumn{4}{c}{$c=6\ (\text{MLP}\ 32:21:15)$}     \\
			\rowcolor[gray]{.8}
			\multicolumn{2}{c}{\cellcolor[gray]{.8}\textbf{Number of BS-user links}}                                                                           & $K=19$                                  & $K=37$                                  & $K=61$                                  & $K=61\ (N_t=128)$           \\
			&Trained with $K=37$ & 121.83 (90.78\%) & 212.89 (92.12\%) &274.53 (88.39\%) & 376.83 (87.12\%)\\
			\multirow{-2}{*}{\textbf{RNN-PGP (POA)}} &  \cellcolor[gray]{.8} Trained with $K \in\{19,\ 37,\ 61\}$                          & \cellcolor[gray]{.8}123.92 (92.34\%)                         & \cellcolor[gray]{.8}210.67 (91.16\%)                         & \cellcolor[gray]{.8}279.31 (89.93\%)                         & \cellcolor[gray]{.8}381.04 (88.09\%) \\
			\bottomrule                   
		\end{tabular}
			}
		\label{table:number of bs-a}
	}
	\label{table:number of bs}
\end{table}

As seen from Fig. \ref{fig: Sum-rate versus runtime}, the RNN-PGP networks have  advantage in terms of the runtime. Specifically, for running 20 iterations, the average runtimes of the ``RNN-PGP (POA)"  ``RNN-PGP (PGP)" are about {\bf 0.0573s} and {\bf 0.0576s}, while the runtimes of the PGP, WMMSE and POA algorithms for 20 iterations are $1.064$s, $1.964$s and $23.231$s, respectively.

%
%
%
%
%
%
%

\begin{figure} [t]
	\centering
	\includegraphics[width=0.4\textwidth]{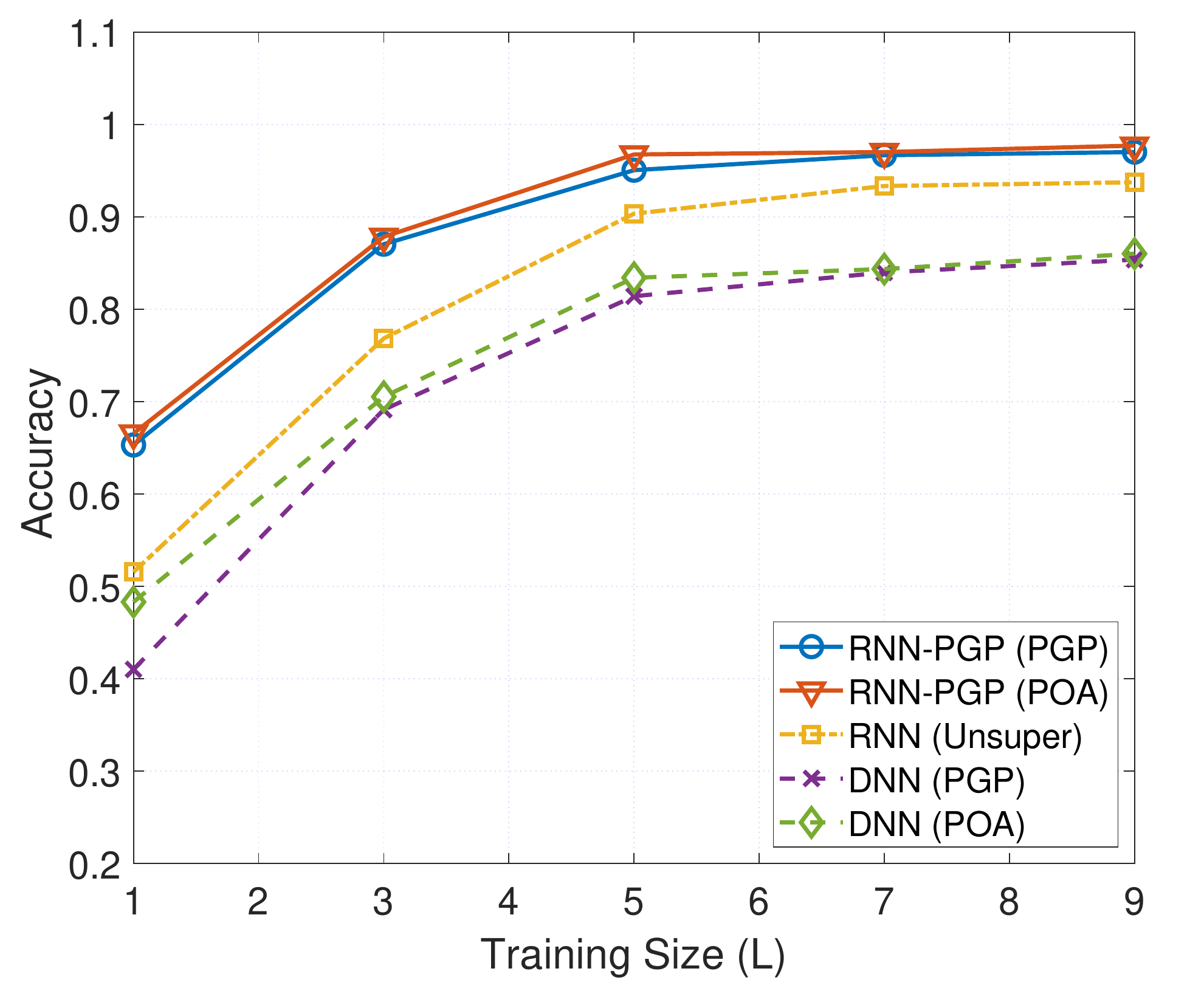}
	\vspace{-0.1cm}
	\caption{\footnotesize Accuracy for different numbers of training samples.}
	\label{fig: accuacyvstrainsize}\vspace{-0.4cm}
\end{figure}

{\bf Impact of training sample size:}
We examine the achieved accuracy versus the size of training data ($L$), as is shown in Fig. \ref{fig: accuacyvstrainsize}. 
From the figure, we can see that all schemes can have improved performance when the number of training samples increases. Moreover, {we compare the performance of RNN-PGP when different training approaches are used. We can see that there is a gap between hybrid training and unsupervised training, but such a gap reduces} when increasing the training data size. This implies that the advantage of hybrid training can be significant if the training size is small. One can also see from the figure that the gap between the black-box based DNN schemes and the proposed RNN-PGP cannot be effectively reduced when the training data size increases.
%

{\bf Generalization w.r.t. number of transmit antennas:} 
To demonstrate the generalization capability of the proposed RNN-PGP, we train the ``RNN-PGP (POA)" using the data set of $N_t=36$ and $K=19$, but test it on data sets with different numbers of $N_t$. The results are shown in the 3rd row of Table \ref{table: number of antennas}.
One can see that the proposed RNN-PGP can yield almost the same accuracy when applied to scenarios with $N_t=36, 72,$ and $108$. Interestingly, we also test the ``RNN-PGP (POA)" in a heterogeneous scenario where the BS can have different numbers of antennas from each other, and the antenna number of each BS is randomly chosen from $16$ to $128$. As seen from the table, the ``RNN-PGP (POA)" can still maintain an average accuracy of $93.78\%$.

In Table \ref{table: number of antennas}, we also present the results when the ``RNN-PGP (POA)" is trained by a mixed data set which contains equal-sized data samples with {$N_t\in \{18, 36, 72, 108\}$}. One can see from the table that this scheme can provide slightly higher accuracy.

%
%


\renewcommand\arraystretch{1.2}
\begin{table}[t]
	\caption{The sum-rate performance of the proposed RNN-PGP for $K=19$ and $N_t =36$.}
	\centering
	\subtable[]{
				\resizebox{\textwidth}{!}{%
		\begin{tabular}{ccccc}
			\toprule
			\bf{Distance Between BSs}                & \multicolumn{4}{c}{$d = 1$ (km)}                                   \\
			\rowcolor[gray]{.8}
			\bf{Number of Neighbors}                 & $c=6$ (MLP 32:21:15)           & $c=6\ (N_t=128)$  & $c=9$ (MLP 45:38:23)          & $c=18$ (MLP 85:73:42)          \\
			\bf{Trained with} $d=1$ (km) & 128.65 (95.86\%) & 156.02 (97.03\%) & 129.31 (96.35\%) & 129.85 (96.75\%) \\
			\rowcolor[gray]{.8} 
			\bf{Trained with $d \in\{0.5,\ 1\}$}            & 127.98 (95.36\%) & 155.86 (96.91\%) & 128.35 (95.63\%) & 128.96 (96.09\%) \\
			\bottomrule
		\end{tabular}
		\label{table:number of neighbors-a}
	}}
	\\
	\subtable[]{        
				\resizebox{\textwidth}{!}{%
		\begin{tabular}{ccccc}
			\toprule
			\bf{Distance Between BSs}                & \multicolumn{4}{c}{$d = 0.5$ (km)} \\
			\rowcolor[gray]{.8}
			\bf{Number of Neighbors}                 &  $c=6$ (MLP 32:21:15)           & $c=6\ (N_t=128)$  & $c=9$ (MLP 45:38:23)           & $c=18$ (MLP 85:73:42)         \\
			\bf{Trained with} $d=1$ (km) &138.15 (82.47\%) & 187.21 (96.95\%) & 146.18 (87.27\%) & 159.22 (95.05\%) \\
			\rowcolor[gray]{.8} 
			\bf{Trained with $d \in\{0.5,\ 1\}$}            & 159.75 (95.37\%) & 187.32 (97.01\%) & 160.16 (95.61\%) & 159.50 (95.22\%)\\
			\bottomrule
	\end{tabular}}
	\label{table:number of neighbors-b}
			}
	\label{table:number of neighbors}
\end{table}

{\bf Generalization w.r.t. number of BSs:}	
To verify the generalization capability of the RNN-PGP with respect to the number of BSs $K$, we consider a training scenario with $N_t=64$, $K=37$ and $c=18$ and $6$, respectively.
From Table \ref{table:number of bs}(a), one can see that with $c=18$, the trained ``RNN-PGP (POA)" can yield a test accuracy of $96.01\%$ for $K=37$, and have slightly reduced accuracies when deployed in scenarios with $K=61$ and $K=91$. We also present in the 4th column of the table the result when the trained RNN-PGP is deployed in a scenario {where} both $K$ and $N_t$ are respectively changed to $91$ and $128$. The achieved accuracy can be maintained around $91.79$\%.

In Table. \ref{table:number of bs}(b), we present another set of results with $c=6$. One can see that the performance degradation is significant when the trained RNN-PGP is deployed to scenarios with different numbers of BSs. This implies that the neighbor size $c$ considered in the RNN-PGP network training should not be too small when compared to $K$.

%
%

{\bf Generalization w.r.t. cell radius:} Here, we examine the generalization capability of the RNN-PGP with respect to the cell radius $d$. We consider a training scenario with {$N_t=36$}, $K=19$ and different number of neighbors $c=6, 9, 18$, and the half inter-BS distance is fixed to $d=1$ km or is mixed with $d\in\{0.5, 1\}$. 

Table \ref{table:number of neighbors}(a) shows the results that the trained frameworks are tested in the scenario with the same cell radius $d=1$ km, while Table \ref{table:number of neighbors}(b) are the results obtained when tested in the scenario with $d=0.5$ km.
By comparing the 3rd rows and first columns of the two tables, one can see that the accuracy decreases from 95.86\% to 82.47\% when the trained RNN-PGP is deployed in a scenario with the cell radius decreased to $0.5$ km, while the sum rate improvement is minor (from 128.65 to 138.15). This implies that the trained network cannot effectively mitigate the inter-cell interference.
Interestingly, as seen from the 2nd columns of the two tables, if we apply the RNN-PGP to the scenario with the antenna number increased to $N_t=128$, then the accuracy degradation due to decreased cell radius becomes minor and the sum rate improvement is more evident (from 156.02 to 187.21).
By comparing the 4th rows and the first two columns of the two tables, one can see that training with mixed cell radius provides good robustness.
Lastly, comparing the first column with the 3rd and 4th columns of Table \ref{table:number of neighbors}(a)-(b), one can also see that larger values of $c$ can make the RNN-PGP to achieve higher accuracy for both $d=1$ km and $d=0.5$ km.

\vspace{-0.3cm}
\subsection{Weighted Sum-rate Performance}
In this part, we consider the MISO interference channel with $N_t = 36$, $K = 19$ and $c=18$.  Each data sample $\ell$ contains $\{\hb_{kj}^{(\ell)}, \alpha_k^{(\ell)}, \wb_k^{(\ell)} \}$, for $\ell=1,\ldots,L$, where the weights $\{ \alpha_k^{(\ell)}\}$ are generated randomly and satisfy $\sum_{k=1}^K\alpha_{k}^{(\ell)} = 1$. 
For the ``RNN-PGP (POA)"  and ``RNN-PGP (PGP)", the sizes of the three hidden layers  are $95, 80,$ and  $45$, respectively. The black-box based ``DNN (POA)" and ``DNN (PGP)" have 5 layers with numbers of nodes equal to
1387 ($2KN_t+K$),  1775, 1775, 1450 and 1368 ($2KN_t$),  respectively.


\begin{figure} [t]
	\centering
	\includegraphics[width=0.4\textwidth]{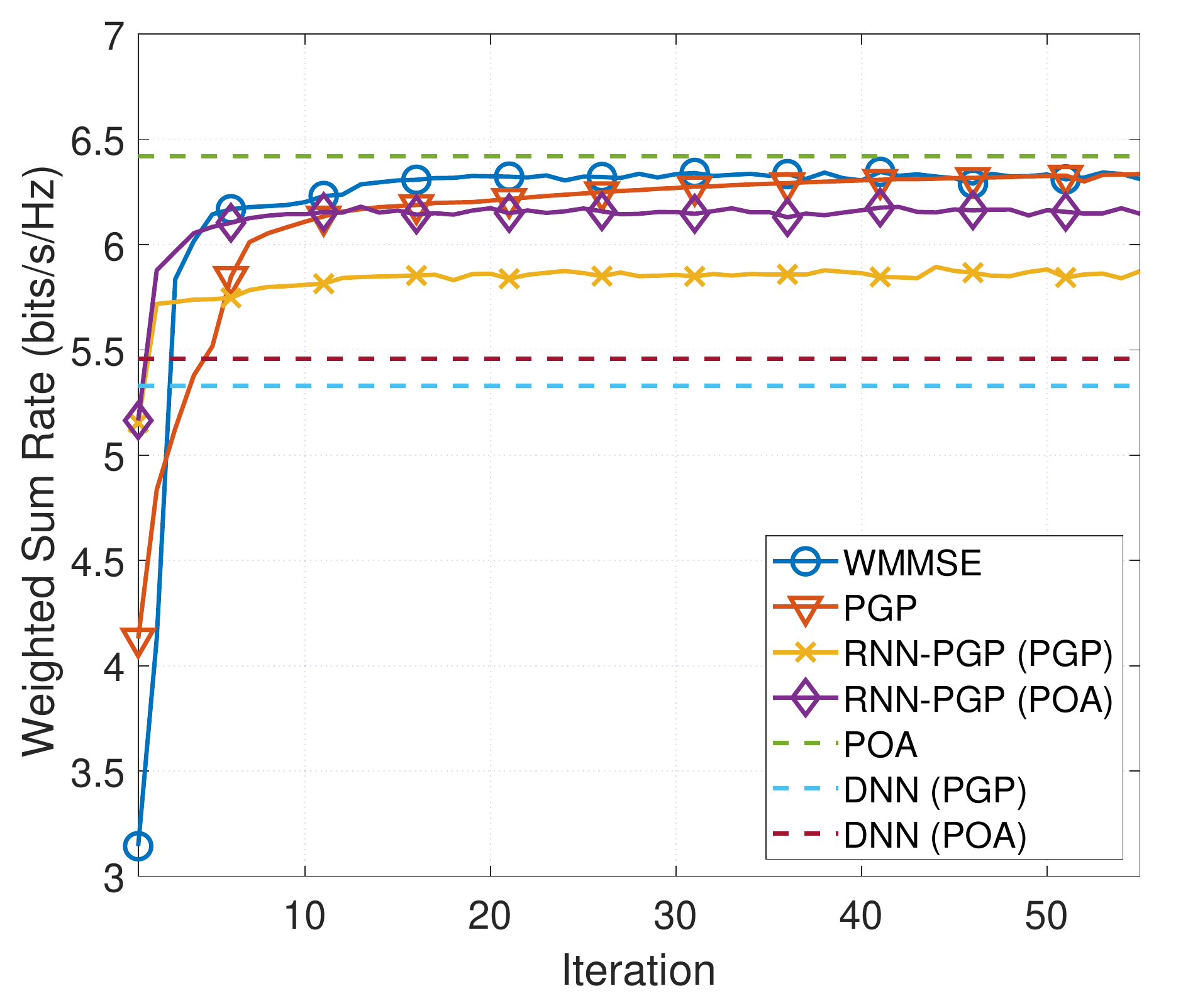}
	\vspace{-0.1cm}
	\caption{\footnotesize WSR versus iteration number for $N_t = 36$, $K = 19$ and $c=18$.}
	\label{fig: Weighted Sum-rate versus iteration}		\vspace{-0.6cm}
\end{figure} 

We can see from Fig. \ref{fig: Weighted Sum-rate versus iteration} that the proposed RNN-PGP can still achieve good WSR performance, especially when trained by the POA solutions.
The black-box based DNNs still suffer poor performance no matter which supervised solutions are used for training.

\begin{table}[t]
	\centering
	\caption{The sum-rate performance of the proposed RNN-PGP under DeepMIMO dataset \cite{alkhateeb2019deepmimo} for $K = 18$ and $c=17$ (MLP 90:75:45).}
	\label{table: number of antennas practical}
	\resizebox{\textwidth}{!}{%
		\begin{tabular}{cccccc}
			\toprule
			\multicolumn{2}{c}{\begin{tabular}[c]{@{}c@{}}\bf{Number of antennas}\\ (number of transmit antennas in $x$, $y$ and $z$ axes)\end{tabular}} & \begin{tabular}[c]{@{}c@{}}$N_t=36$\\ $(X=1, Y=6, Z=6)$\end{tabular} & \begin{tabular}[c]{@{}c@{}}$N_t=72$\\ $(X=1, Y=6, Z=12)$\end{tabular} & \begin{tabular}[c]{@{}c@{}}$N_t=108$\\ $(X=3, Y=6, Z=6)$\end{tabular}  & \begin{tabular}[c]{@{}c@{}}Random number antennas selected from\\ $(X\cdot Y\cdot Z \in[16,128])$\end{tabular}\\
			\rowcolor[gray]{.8} 
			\multicolumn{2}{c}{\cellcolor[gray]{.8}\textbf{PGP}}                                            & $102.04$                                   & 110.36                                  & 114.94                                   & -                                   \\
			& Trained by POA with $X=1, Y=6, Z=6$                              & $98.49\ (96.53\%)$                         & $104.15\ (94.37\%)$                        & $ 108.18\ (94.12\%)$                         &- $(93.57\%)$                          \\
			\multirow{-2}{*}{\textbf{RNN-PGP}} & \cellcolor[gray]{.8}Trained with $N_t \in\{36,\ 72,\ 108\}$ & \cellcolor[gray]{.8} $97.38\ (95.43\%)$ & \cellcolor[gray]{.8}$105.05\ (95.19\%)$ & \cellcolor[gray]{.8}$109.22\ (95.02\%)$ & \cellcolor[gray]{.8}- $(95.09\%)$ \\
			\bottomrule              
		\end{tabular}%
	}
\end{table}

\vspace{-0.3cm}
\subsection{Performance on DeepMIMO dataset}

\begin{figure} [t]
	\centering
	\includegraphics[width=0.55\textwidth]{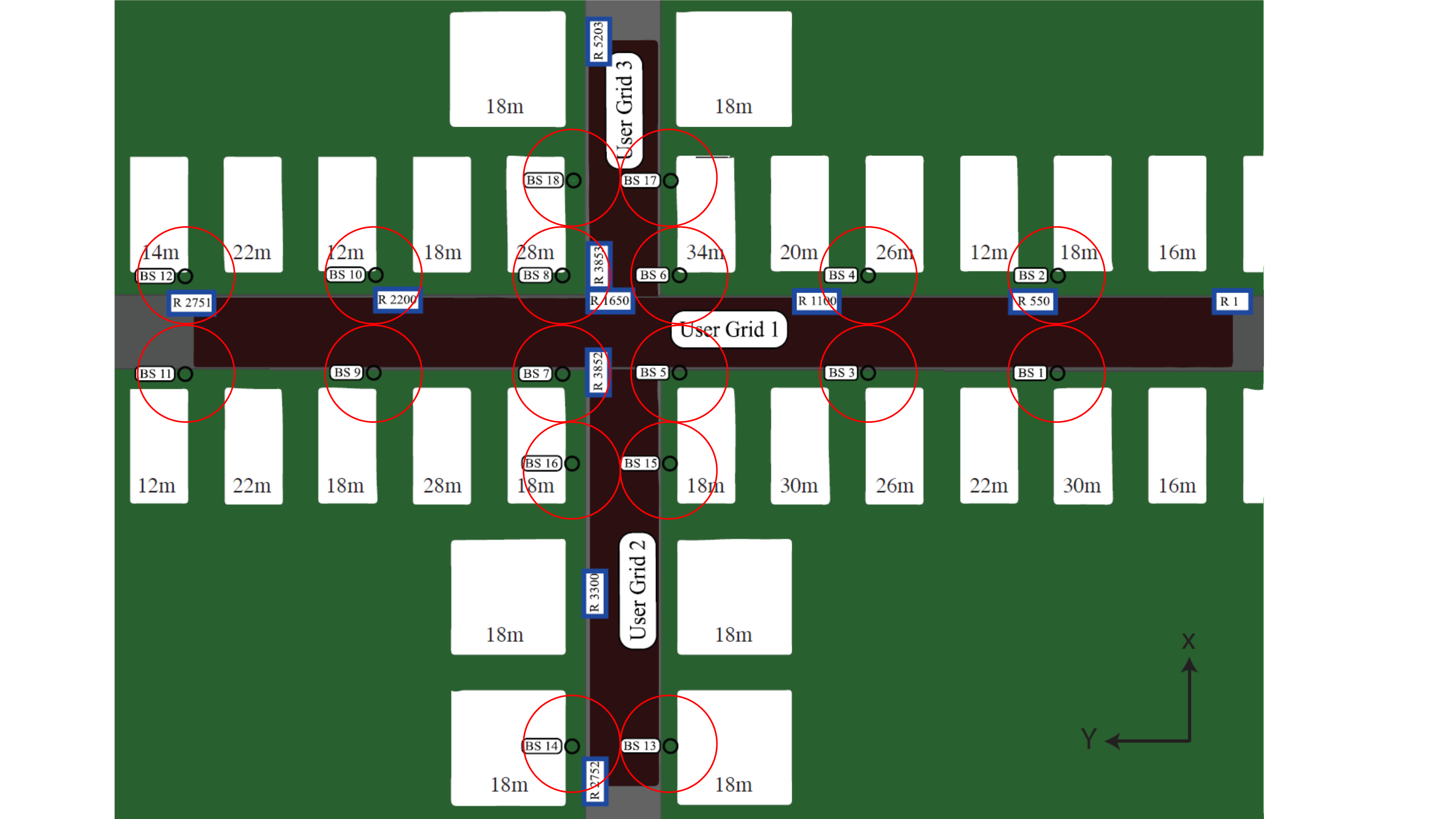}
	\vspace{-0.1cm}
	\caption{\footnotesize The outdoor scenario provided in 'O1' of \cite{alkhateeb2019deepmimo}}
	\label{fig: open channel}	\vspace{-0.4cm}
\end{figure}

In this subsection, we test the proposed RNN-PGP on the ray-tracing based DeepMIMO dataset \cite{alkhateeb2019deepmimo}. 
We consider an outdoor scenario `O1’ 
as shown in Fig. \ref{fig: open channel}. 
The main street (the horizontal one) is $600$m long and $40$m wide, and the second street (the vertical one) is $440$m long and $40$m wide.  We consider 18 BSs ($K= 18$), and their served UEs are selected randomly around their respective BS within $50$m on the streets (the intersection of the black street and the red circle in Fig. \ref{fig: open channel}). 
The BSs are equipped with antennas with dimensions $X, Y,$ and $Z$ along the x, y, and z axes, and the antenna size is $N_t=XYZ$.
Analogous to the synthetic data, we generated 5000 training samples $(L=5000)$ and 1000 test samples. 

\begin{figure} [t]
	\centering
	\includegraphics[scale=0.4]{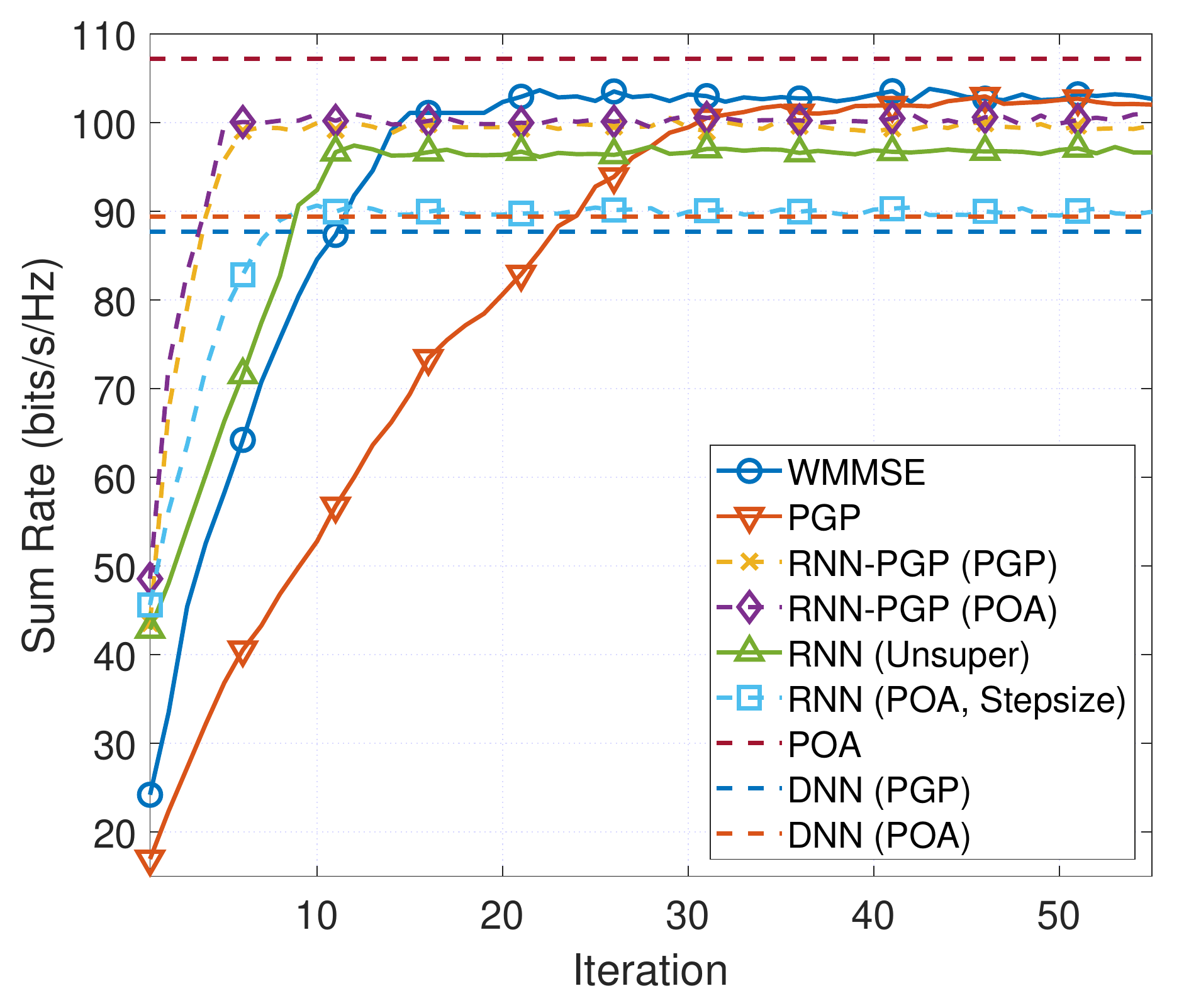}
	\vspace{-0.1cm}
	\caption{\footnotesize Sum-rate versus iteration in DeepMIMO channel for $N_t = 36\ (X =1, Y= 6, Z=6)$, $K=18$ and $c=17$.}
	\label{fig: Sum-rate versus iteration in practice channel}\vspace{-0.6cm}
\end{figure}

In this experiment, for the proposed RNN-PGP, the numbers of neurons of the 5-layer MLP are $72, 90, 75, 45$ and $37$, respectively; the black-box based DNNs also have 5 layers with numbers of neurons equal to $2592, 2845, 2450, 1450$ and $1296$, respectively. The testing results versus the iteration number are shown in Fig. \ref{fig: Sum-rate versus iteration in practice channel}. Again, we can observe consistent results and the proposed RNN-PGP can  achieve good sum rate performance.

Similar to Table \ref{table: number of antennas}, we test the generalization capability of the RNN-PGP in the DeepMIMO data set. As seen from Table \ref{table: number of antennas practical}, the proposed RNN-PGP still can be generalized well and maintain good accuraces. 

\renewcommand\arraystretch{1.44}
\begin{table*}[t]
	\caption{The sum-rate performance of the proposed RNN-PGP under DeepMIMO dataset \cite{alkhateeb2019deepmimo} for $K_r=K_t$.}
	\centering
	\subtable[]{
					\resizebox{\textwidth}{!}{%
		\begin{tabular}{ccccc}
			\toprule
			\multicolumn{2}{c}{\begin{tabular}[c]{@{}c@{}}\bf{Number of BSs}\\ \{active\ BSs\}\end{tabular}}                                                                & \multicolumn{3}{c}{\begin{tabular}[c]{@{}c@{}}$K_t = K_r = 6$ (MLP 45:32:28)\\ $\{5,6,7,8,15,16\}$\end{tabular}}  \\
			\rowcolor[gray]{.8}
			\multicolumn{2}{c}{\cellcolor[gray]{.8}\begin{tabular}[c]{@{}c@{}}\bf{Number of antennas}\\ (number of transmit antennas in $x$, $y$ and $z$ axes)\end{tabular}} &\cellcolor[gray]{.8} \begin{tabular}[c]{@{}c@{}}$N_t=36$\\ $(X=1, Y=6, Z=6)$\end{tabular} &\cellcolor[gray]{.8} \begin{tabular}[c]{@{}c@{}}$N_t=64$\\ $(X=1, Y=8, Z=8)$\end{tabular} &\cellcolor[gray]{.8} \begin{tabular}[c]{@{}c@{}}$N_t=216$\\ $(X=6, Y=6, Z=6)$\end{tabular} \\
			& Trained with $X=1, Y=8, Z=8$                                & $24.56\ (92.49\%)$                                                                    & $28.12\ (94.56\%)$                                                                  & $36.07\ (92.47\%)$                                                                \\
			\multirow{-2}{*}{\textbf{\begin{tabular}[c]{@{}c@{}}RNN-PGP\\ (Trained by PGP)\end{tabular}}}      & \cellcolor[gray]{.8}Trained with $N_t \in \{36,\ 64,\ 216\}$     & \cellcolor[gray]{.8}$24.84\ (93.52\%)$                                            & \cellcolor[gray]{.8}$27.80\ (93.49\%)$                                            & \cellcolor[gray]{.8}$36.47\ (93.51\%)$                                              \\
			\bottomrule                                    
		\end{tabular}   
				}
	}
	\\
	\subtable[]{        
						\resizebox{\textwidth}{!}{%
		\begin{tabular}{ccccc}
			\toprule
			\multicolumn{2}{c}{\begin{tabular}[c]{@{}c@{}}\bf{Number of BSs}\\ \{active\ BSs\}\end{tabular}}                                                                 & \multicolumn{3}{c}{\begin{tabular}[c]{@{}c@{}}$K_t = K_r = 12$ (MLP 75:56:50)\\ $\{3,4,5,6,7,8,9,10,15,16,17,18\}$\end{tabular}}                                                                                                            \\
			\rowcolor[gray]{.8}
			\multicolumn{2}{c}{\cellcolor[gray]{.8}\begin{tabular}[c]{@{}c@{}}\bf{Number of antennas}\\ (number of transmit antennas in $x$, $y$ and $z$ axes)\end{tabular}} &\cellcolor[gray]{.8} \cellcolor[gray]{.8} \begin{tabular}[c]{@{}c@{}}$N_t=36$\\ $(X=1, Y=6 ,Z=6)$\end{tabular} &\cellcolor[gray]{.8} \begin{tabular}[c]{@{}c@{}}$N_t=64$\\ $(X=1, Y=8, Z=8)$\end{tabular} &\cellcolor[gray]{.8} \begin{tabular}[c]{@{}c@{}}$N_t=216$\\ $(X=6, Y=6,Z=6)$\end{tabular} \\
			& Trained with $X=1, Y=8, Z=8$                                & $37.46\ (92.44\%)$                                                                    & $43.72\ (94.23\%) $                                                                   & $54.99\ (92.40\%) $                                                                     \\
			\multirow{-2}{*}{\textbf{\begin{tabular}[c]{@{}c@{}}RNN-PGP\\ (Trained by PGP)\end{tabular}}}      & \cellcolor[gray]{.8}Trained with $N_t \in\{36,\ 64,\ 216\}$     & \cellcolor[gray]{.8}$37.88\ (93.49\%) $                                             &  \cellcolor[gray]{.8}$43.35\ (93.43\%) $                                             & \cellcolor[gray]{.8}$55.42\ (93.13\%) $        \\
			\bottomrule                                    
	\end{tabular}}
			}
	\label{table:sum-rate coop multi-cell}
\end{table*}

\begin{table*}[t]
	\centering
	\caption{The sum-rate performance of the proposed RNN-PGP under DeepMIMO dataset \cite{alkhateeb2019deepmimo} for $N_t = 12$, $K_r = 12$ and $N_t=64\ (X=1, Y= 8, Z=8)$, (MLP 75:56:50).}
	\label{table: number of cells}
		\resizebox{\textwidth}{!}{%
	\begin{tabular}{cccccc}
		\toprule
		\multicolumn{2}{c}{\textbf{Number of BSs} $K_t$}                                                                                                                            & $K_t=6$                                 & $K_t=12$                               & $K_t =18$                               & Random selected from 6-18           \\
		\rowcolor[gray]{.8} 
		\multicolumn{2}{c}{\cellcolor[gray]{.8}\textbf{PGP}}                                                                                             & 39.47                                  & 46.45                                 & 69.66                                 & -                                   \\
		& Trained with $K_t=12$                             & 36.57 (92.36\%)                         & 43.76 (94.23\%)                        & 64.21 (92.17\%)                         & - (92.78\%)                         \\
		\multirow{-2}{*}{\textbf{\begin{tabular}[c]{@{}c@{}}RNN-PGP\\ (Trained by PGP)\end{tabular}}} & \cellcolor[gray]{.8}Trained with $K_t \in\{6,\ 12,\ 18\}$ & \cellcolor[gray]{.8}36.88 (93.43\%) & \cellcolor[gray]{.8}43.6 (93.89\%) & \cellcolor[gray]{.8}64.96 (93.26\%) & \cellcolor[gray]{.8}- (93.31\%)\\
		\bottomrule
	\end{tabular}%
		}
\end{table*} 
\vspace{-0.3cm}
\subsection{Performance for Cooperative Multicell Beamforming}

In this subsection, we examine the proposed RNN-PGP in Fig. \ref{fig:coop-rnn-network} for the cooperative multicell beamforming problem in Section \ref{sec:other applications}. 
We again consider the outdoor scenario provided in 'O1' of the DeepMIMO dataset \cite{alkhateeb2019deepmimo}. Two cases are considered: (1) $K_t=K_r=6$, and (2) $K_t=K_r=12$.
In case (1), the BSs 5, 6, 7, 8, 15, 16 in Fig. \ref{fig: open channel} are selected, while BSs 3, 4, 5, 6, 7, 8, 9, 10, 15, 16, 17, 18 in Fig. \ref{fig: open channel} are selected in case (2). 
The UEs are selected randomly on the streets (the black region of Fig. \ref{fig: open channel}). For both cases, the RNN-PGP is trained by the dataset with $N_t=64\ (X=1, Y=Z=8)$ and the PGP solutions.

The test performance results for the two cases are shown in Table \ref{table:sum-rate coop multi-cell}(a) and Table \ref{table:sum-rate coop multi-cell}(b), respectively. One can observe that for both cases the proposed RNN-PGP can yield high accuracy and generalizes well when deployed in scenarios with different number of transmit antennas. 

In Table. \ref{table: number of cells}, we further consider the generalization capability with respect to the number of BSs $K_t$. The RNN-PGP is trained under the setting of case (2) with $K_t=12$, $K_r=12$ and $N_t=64$ while is tested in different scenarios with $K_t=6$, $K_t=18$ and randomly selected numbers of BSs. It can be observed from the table that the proposed RNN-PGP can maintain good performance and has good generalization capability w.r.t the number of BSs. 

%
\vspace{-0.3cm}
\section{Conclusion}\label{sec: conclusion}

In this paper, we have considered a learning-based beamforming design for MISO interference channels and cooperative multicell scenarios. In particular, in order to overcome the computational issues of massive MIMO beamforming optimization, we have proposed the RNN-PGP network by unfolding the simple PGP method. We have shown that by exploiting the low-dimensional structures of optimal beamforming solutions as well as the gradient vector of the WSR, the {proposed} RNN-PGP network can have a low complexity, {and such complexity} is independent of the number of transmit antennas. We have also refined the input and output of the MLP in the RNN-PGP to remove its dependence on the number of BSs.
Extensive experiments have been conducted based on both synthetic channel dataset and the DeepMIMO dataset. The presented experiment results have demonstrated that the proposed RNN-PGP can achieve a high solution accuracy with the expense of a small computation time. More importantly, the proposed RNN-PGP has promising generalization capabilities with respect to the number of transmit antennas, the number of BSs, and the cell radius, which is a key ability to be employed in heterogeneous networks.

\footnotesize

\end{document}